\documentclass[a4paper,12pt]{article}
\pdfoutput =1 

\usepackage{jheppub} 

\usepackage[T1]{fontenc} 
\usepackage{float} 
\usepackage{amsmath}
\usepackage{graphicx}
\usepackage{slashed}
\usepackage{amsfonts}   
\usepackage{amssymb}
\usepackage[capitalise]{cleveref}

\title{\boldmath SYK/AdS duality with Yang-Baxter deformations}

\author[a]{Arindam Lala,}
\author[b]{and Dibakar Roychowdhury}

\affiliation[a]{Instituto de F\'{i}sica, Pontificia Universidad Cat\'{o}lica de Valpara\'{i}so,\\ Casilla 4059, Valparaiso, Chile}
\affiliation[b]{Department of Physics, Swansea University,\\Singleton Park, Swansea SA2 8PP, United Kingdom}

\emailAdd{arindam.lala@pucv.cl}
\emailAdd{dibakarphys@gmail.com, Dibakar.RoyChowdhury@swansea.ac.uk}

\abstract{In this paper, based on the notion of SYK/AdS duality we explore the effects of Yang-Baxter (YB) deformations on the SYK spectrum at strong coupling. In the first part of our analysis, we explore the consequences of YB deformations through the Kaluza-Klein (KK) reduction on $ (AdS_2)_{\eta}\times (S^1)/Z_2 $. It turns out that the YB effects (on the SYK spectrum) starts showing off at \textit{quadratic} order in $ 1/J $ expansion. For the rest of the analysis, we provide an interpretation for the YB deformations in terms of bi-local/collective field excitations of the SYK model. Using large $ N $ techniques, we evaluate the effective action upto quadratic order in the fluctuations and estimate $ 1/J^2 $ corrections to the correlation function at strong coupling.}
\begin{document} 
\maketitle
\flushbottom
\section{Overview and Motivation}
Very recently, the SYK model \cite{Sachdev:1992fk}-\cite{Forste:2017apw} has been proposed as one of those handful examples (within the realm of AdS/CFT correspondence) where one might hope to solve the spectrum associated with the quantum mechanical system at strong coupling\footnote{See \cite{Rosenhaus:2018dtp} for a nice comprehensive review.}. Other than its solvability, this $ (1+0) $D system of ($ N\gg1 $) fermions possesses two other remarkable features namely,-(I) the Lyapunov exponent associated with the out of time ordered four point correlators saturates the bound \cite{Shenker:2013pqa}-\cite{Maldacena:2015waa} and (II) the emerging scale invariance at low energies.\\
Understanding the dual gravitational counterpart \cite{Teitelboim:1983ux}-\cite{Engelsoy:2016xyb} corresponding to the SYK model had always been challenging until very recently \cite{Das:2017pif}-\cite{Das:2017wae}. In \cite{Das:2017pif}, the authors provide a very solid evidence in favour of the dual 3D gravitational counterpart corresponding to the zero temperature version of the model where they explore the Kaluza-Klein (KK) tower associated with the scalar field excitations on $ AdS_2 \times (S^1)/Z_2 $. In the strict IR ($ |J t |\gg 1 $) limit, the metric along the compact third direction becomes constant, whereas on the other hand, away from the fixed point it acquires non trivial dependence on the $ AdS_2 $ coordinates through the dilaton profile. In their analysis \cite{Das:2017pif}, the authors explore the $ 1/J $ corrections to the zero modes of the theory and compute the strong coupling corrections to the propagator that precisely matches to that with the earlier results in \cite{Maldacena:2016hyu}.\\
The purpose of the present paper is to explore and understand this duality conjecture in the presence of so called Yang-Baxter (YB) deformations \cite{Kyono:2017jtc}-\cite{Okumura:2018xbh} and also to understand its possible consequences on the corresponding bi-local excitations \cite{Jevicki:2016bwu}-\cite{Jevicki:2016ito} associated with the SYK model at strong coupling. The motivation behind our analysis strictly follows from holography where we start with the deformed $ AdS_2 $ version of the theory in the bulk \cite{Kyono:2017jtc} and lift it to three dimensions in order to compute the KK spectrum associated with the scalar field excitations in the dual gravitational counterpart. Our analysis reveals that the holographic correspondence between the SYK model and its dual gravitational counterpart brings into a non trivial $ 1/J^2 $ corrections to the spectrum of the SYK model which has its origin in the YB deformations associated with the $ (AdS_2)_{\eta}\times (S^1)/Z_2 $ theory in ($ 2+1 $) D. For the rest of our analysis, we search for an interpretation of such deformations in terms of \textit{collective field } excitations \cite{Jevicki:1980zg} associated with the SYK model. Looking at the holographic side of the duality, we propose a possible YB \textit{scaling} of the collective excitations within the SYK model and compute the effective action at quadratic order in the fluctuations (around IR critical point) and thereby the corresponding propagator at strong coupling ($ |Jt|\gg 1 $).  Our analysis reveals that the effective action receives non trivial $ 1/J^2 $ corrections that has remarkable structural similarity to that with the corresponding quadratic action associated with scalar field excitations computed on the dual gravitational counterpart of the theory. \\
The organisation of the paper is as follows: In section 2, we briefly review the YB deformations and its implications on the Almheiri-Polchinski (AP) model \cite{Kyono:2017jtc}-\cite{Okumura:2018xbh}. In section 3, we compute the zero modes associated with the spectrum and estimate the $ 1/J^2 $ corrections to it. In section 4, we provide a possible interpretation of the YB effects on the collective excitations within the SYK model. Finally, we conclude in section 5. 
\section{YB deformations and string theory}
\subsection{Basics}
The primary motivation behind introducing the Yang-Baxter (YB) deformations (associated with non-linear $\sigma$-models, e.g; the Principal Chiral Model (PCM)) was the observation 
that the later is equivalent to a two-dimensional field theory (defined on some compact manifold $\mathcal{M}$) embodied with a rank 2 symmetric tensor (metric) field ($ \gamma^{\alpha \beta} $) as well as an antisymmetric two-form. These YB $\sigma$-models are characterized by 
some $\mathbb{R}$-linear operators \cite{Klimcik:2002,Delduc:2013,Yoshida:2015a} and seem to posses a left-symmetry together with the Poisson-Lie symmetry with respect to the right action of the group on itself. These symmetries of the model could be associated with certain types of dualities embedded in its structure\footnote{For more details, the interested reader is encouraged to go through the references \cite{Klimcik:2002}. }. For any generic Lie group $\mathcal{G}$ with Lie algebra $\textsl{g}=Lie(\mathcal{G})$, the action corresponding to YB $\sigma$-models could be formally expressed as\footnote{The $ \mathfrak{R} $ operator satisfies the so called Yang-Baxter (YB) equation (\ref{2.6}) which we elaborate in the next secion in the context of deformations associated with the $ AdS_2 $ supercosets.} \cite{Klimcik:2002,Delduc:2013,Yoshida:2015a} ,
\begin{equation}
\label{act.ybs}
S=-\frac{1}{2}\left(\gamma^{\alpha\beta}-\epsilon^{\alpha\beta}\right) \int_{-\infty}^{\infty} d\tau \int_{0}^{2\pi}d\sigma
\; \text{Tr}\left(J_{\alpha}\frac{1}{1-\eta~ \mathfrak{R}_{\mathfrak{g}}}J_{\beta}\right)
\end{equation}
where $(\tau,\sigma)$ are the two-dimensional world-sheet coordinates together with the skew-symmetric tensor $\epsilon^{\alpha\beta}$ normalized as, $\epsilon^{\tau\sigma}=-\epsilon^{\sigma\tau}=1$. Here, 
$J_{\alpha}=\mathfrak{g}^{-1}\partial_{\alpha}\mathfrak{g}$ is the left-invariant one-form expressed in terms of $\mathfrak{g}(\tau,\sigma)
\in \mathcal{G}$ and the trace is defined over the fundamental representation of the algebra $\textsl{g}$. It is indeed trivial to notice that in the limit of the vanishing ($\eta\rightarrow 0$) YB deformation one recovers the sigma model corresponding to that of the PCM. As an additional fact, the YB deformed version of an integrable sigma model seems the preserve the integrable structure as well\footnote{The $ AdS_2 $ supercoset model (that is realized as a solution in the framework of 2D dilaton gravity \cite{Teitelboim:1983ux}-\cite{Engelsoy:2016xyb}) does not preserve integrability as the dual SYK turns out to be maximally \textit{chaotic}. As a consequence of this, the YB deformed version of it is not expected to be integrable as well.}.\\
The motivation behind introducing the YB deformations in the context of string sigma models stems from the fact that they play crucial role towards a profound understanding of the underlying dynamics in AdS/CFT correspondence. It eventually includes a broader classes of stringy geometries \cite{Bakhmatov:2018apn} within the unified framework of gauge/string duality. Referring back to the original Maldacena duality between type IIB super-strings propagating in $ AdS_5 \times S^5 $ and that of $ \mathcal{N}=4 $ SYM in 4D, the YB deformations applied to the corresponding string sigma model provide a non trivial \textit{generalization} of the duality. On the gauge theory side, these deformations could be realized as a $ q $ deformation of the (say for example $ SO(6) $) spin chain \cite{Beisert:2012} which thereby preserves integrability like in the usual $ \mathcal{N}=4 $ SYM \cite{Minahan:2002}. On the other hand, on the gravity side one could generate a wider class of dual geometries depending on different types of solutions associated with the classical $ \mathfrak{R} $ matrix \cite{Yoshida:2014}-\cite{Delduc:2013a}.\\
Keeping the spirit of the above discussion, the purpose of the present paper is to \textit{generalize} the notion of $ AdS_2/SYK $ duality \cite{Das:2017pif} in the presence of YB deformations. Very recently, the YB deformation of the Almheiri-Polchinski
model has been constructed in \cite{Kyono:2017jtc} and the dual (deformed) SYK version of this gravity model is yet to be constructed.  The purpose of the present analysis is to fill up this gap and provide a systematic realization of the dual gauge theory at strong coupling.
\subsection{The deformed AP model}\label{Def.AP}
The purpose of this Section is to provide a brief introduction to the Almheiri-Polchinski (AP) model \cite{Almheiri:2014cka} and its Yang-Baxter (YB) form of deformations introduced very recently in \cite{Kyono:2017jtc}. The Yang-Baxter (YB) deformation of the $ AdS_2 $ metric is based on the usual notion of coset space formulation of the 2D non-linear sigma model that acts both on the metric as well as the anti-symmetric two form field. For the metric part, the YB deformation was introduced as \cite{Kyono:2017jtc},
\begin{eqnarray}
ds_{\eta}^2 = 2 Tr\left[ J \frac{1}{1-2 \eta \mathfrak{R}_{\mathfrak{g}}\circ P}P(J)\right] \label{E1}
\end{eqnarray}
where, $ \eta $ is the deformation parameter. Here, the left invariant one form $ J(\in \mathfrak{sl}(2)) $ is defined as usual,
\begin{eqnarray}
J= \mathfrak{g}^{-1}d\mathfrak{g}
\end{eqnarray}
where $ \mathfrak{g} $ is an element of $ SL(2)/U(1) $. 
The projection could be defined as follows\cite{Kyono:2017jtc},
\begin{eqnarray}
P(X)=\frac{Tr(X T_0)}{Tr(T_0 T_0)}T_0 +\frac{Tr(X T_1)}{Tr(T_1 T_1)}T_1,~~X\in \mathfrak{sl}(2)
\end{eqnarray}
together with the chain operation of the following form,
\begin{eqnarray}
\mathfrak{R}_{\mathfrak{g}}(X)=\mathfrak{g}^{-1}\circ \mathfrak{R}(\mathfrak{g}X \mathfrak{g}^{-1})\circ \mathfrak{g}
\end{eqnarray}
where the linear operator $ \mathfrak{R} $ satisfies the modified classical YB (mCYBE) equation of the following form,
\begin{eqnarray}
[\mathfrak{R}(X),\mathfrak{R}(Y)]-\mathfrak{R}([\mathfrak{R}(X),Y]+[X,\mathfrak{R}(Y)])=c.[X,Y]\label{2.6}
\end{eqnarray}
where, $ c\neq 0 $ for mCYBE and $ c=0 $ for \textit{homogeneous} CYBE \cite{Kyono:2017jtc}.
Taking the generators, $ T_I \in \mathfrak{so}(1,2) $ in the fundamental representation,
\begin{eqnarray}
T_I \sim \sigma_I
\end{eqnarray}
(where, $ \sigma_I $s are the Pauli matrices) a straightforward calculation of (\ref{E1}) reveals \cite{Kyono:2017jtc},
\begin{eqnarray}
ds_{\eta}^2 = \mathcal{F}_{\eta}(X.P)\frac{-dt^2 +dz^2}{z^2}\label{E7}
\end{eqnarray}
where, the function $  \mathcal{F}_{\eta}(X.P) $ could be formally expressed as,
\begin{eqnarray}
\mathcal{F}_{\eta}(X.P)&=&\frac{1}{1-\eta^2 (X.P)^2}\nonumber\\
X.P &=& \frac{1}{z}(\alpha +\beta t +\gamma (-t^2+z^2)).
\end{eqnarray}
Here, $ \alpha $, $\beta $ and $ \gamma $ are three constant parameters of the theory that satisfy the following constraint,
\begin{eqnarray}
\beta^2 +4 \alpha \gamma + 4c=0.
\end{eqnarray}
Clearly in the limit, $ \eta \rightarrow 0 $ one recovers the usual $ AdS_2 $ metric as a part of the full background solutions corresponding to the AP model \cite{Almheiri:2014cka}.
The major accomplishment of \cite{Kyono:2017jtc} was to embed the above YB deformed $ AdS_2 $ metric (\ref{E7}) as a solution within $ 1+1 $ D dilaton gravity system\footnote{In their analysis, the authors set the matter part of the Lagrangian equal to zero \cite{Kyono:2017jtc}.},
\begin{eqnarray}
S^{(\eta)}_{g}=\frac{1}{16 \pi G}\int d^2x \sqrt{-g}(\Phi_{\eta}^{2}R-U(\Phi^{2}_{\eta}))\label{E10}
\end{eqnarray}
with a particular type of $ \eta $ modification introduced to the dilaton potential ($ U(\Phi) $) that drives the potential from its standard quadratic form to a hyperbolic function. The resulting dilaton profile could be formally expressed as \cite{Kyono:2017jtc},
\begin{eqnarray}
\Phi_{\eta}^{2}=\frac{1}{2 \eta}\log \left[\frac{1+ \eta (X.P)}{1-\eta (X.P)} \right]+1. 
\end{eqnarray}
\section{3D holography}
In this section we intend to build up a notion for 3D holography in the presence of YB deformations described above. The first step towards this direction would be to uplift the deformed AP model (\ref{E10}) in one higher dimension by introducing an additional compact direction ($ \Theta $) and compute the corresponding scalar spectrum associated with Kaluza-Klein (KK) modes \cite{Das:2017pif}. 
\subsection{A 3D uplift}
We propose the 3D metric of the following form,
\begin{eqnarray}
d\mathsf{S}_{\eta}^2 = \mathsf{G}_{MN}dX^{M}dX^{N}=\mathcal{F}_{\eta}(X.P)\frac{-dt^2 +dz^2}{z^2}+\mathsf{G}^{(\eta)}_{\Theta\Theta}d\Theta^{2}.\label{E12}
\end{eqnarray}
A straightforward computation reveals,
\begin{eqnarray}
\frac{1}{16 \pi G^{(3)}}\int d^3 x\sqrt{-\mathsf{G}} \mathsf{R}^{(3)}&=&\frac{\Sigma_{I}}{16 \pi G^{(3)}}\int d^2x\sqrt{-g}\sqrt{\mathsf{G}^{(\eta)}_{\Theta\Theta}}\left( R +\frac{z^2 \mathcal{I}^{(\eta)}}{4 \mathcal{F}_{\eta}(X.P)\mathsf{G}^{(\eta)2}_{\Theta\Theta}}\right)\nonumber\\
&=& \frac{\Sigma_{I}}{16 \pi G^{(3)}}\int d^2x\sqrt{-g}\sqrt{\mathsf{G}^{(\eta)}_{\Theta\Theta}}\left( R +\varphi_{\eta}\right)
\end{eqnarray}
where\footnote{Here, $ \square^{(2)}=-\partial^{2}_{t}+\partial^{2}_{z} $ is the 2D Laplacian.},
\begin{eqnarray}
\mathcal{I}^{(\eta)}&=& (\partial_t \mathsf{G}^{(\eta)}_{\Theta \Theta})^{2}-(\partial_z \mathsf{G}^{(\eta)}_{\Theta \Theta})^{2}+2\mathsf{G}^{(\eta)}_{\Theta\Theta}(-\partial^{2}_{t}+\partial_z^{2})\mathsf{G}^{(\eta)}_{\Theta\Theta}\nonumber\\
&=& (\partial_t \mathsf{G}^{(\eta)}_{\Theta \Theta})^{2}-(\partial_z \mathsf{G}^{(\eta)}_{\Theta \Theta})^{2}+2\mathsf{G}^{(\eta)}_{\Theta\Theta}\square^{(2)}\mathsf{G}^{(\eta)}_{\Theta\Theta}
\end{eqnarray}
and $ \Sigma_{I} $ is the corresponding volume of the compact manifold.
As a trivial check of our analysis, we first notice that in the limit, $ \eta \rightarrow 0 $  and with the following choice of the metric,
\begin{eqnarray}
\mathsf{G}^{(\eta =0)}_{\Theta\Theta}=\left(\frac{\alpha +\gamma  \left(z^2-t^2\right)+\beta  t}{z}+1\right)^2
\end{eqnarray}
one correctly reproduces the desired  form of the dilaton potential \cite{Kyono:2017jtc},
\begin{eqnarray}
\sqrt{\mathsf{G}^{(\eta =0)}_{\Theta\Theta}}\varphi_{\eta =0}=-U(\Phi^{2}_{\eta =0} )=2(\Phi^{2}_{\eta =0}-1).
\end{eqnarray}
The above scenario generalizes quite non-trivially for non zero deformations,
\begin{eqnarray}
\mathsf{G}^{(\eta)}_{\Theta\Theta}=\left(\frac{\log \left(\frac{1+\eta  (X.P)}{1-\eta (X.P)}\right)}{2 \eta }+1\right)^2
\end{eqnarray}
that reproduces the desired form of the dilaton potential \cite{Kyono:2017jtc},
\begin{eqnarray}
\sqrt{\mathsf{G}^{(\eta )}_{\Theta\Theta}}\varphi_{\eta}=-U(\Phi^{2}_{\eta } ).
\end{eqnarray}
Therefore like in the undeformed scenario, the YB deformed ($ 1+1 $) D dilaton gravity could be uplifted to $ (2+1) $ D with dilaton being the third direction. The above background (\ref{E12}) would serve as the starting point of our subsequent analysis.
\subsection{Kaluza-Klein modes}\label{Sec.KK}
The purpose of this Section is to obtain the KK spectrum of a single scalar field ($ \phi $) over the $ \eta $ deformed background (\ref{E12}) obtained previously. We start with the scalar action of the following form,
\begin{eqnarray}
S_{\phi}=\frac{1}{2}\int d^3 x\sqrt{-\mathsf{G}}\left[ -\mathsf{G}^{MN}\partial_M \phi \partial_N \phi - m^{2}\phi^2 - V(\Theta)\phi^2\right] 
\end{eqnarray}
where, $ V(\Theta)=V\delta (\Theta) $ is the delta function potential as usual \cite{Das:2017pif}.
In the subsequent analysis we focus on the case with \textit{homogeneous} CYBE  which amounts of setting \cite{Kyono:2017jtc},
\begin{eqnarray}
X.P = \frac{\alpha}{z},~~X^2=-1,~~P^2 =0
\end{eqnarray}
and simplifies the background metric (\ref{E12}) as,
\begin{align}
\begin{split}
d\mathsf{S}_{\eta}^2 &= \mathsf{G}_{MN}dX^{M}dX^{N}=\mathcal{F}_{\eta}(z)
\frac{-dt^2 +dz^2}{z^2}+\mathsf{G}^{(\eta)}_{\Theta\Theta}(z)d\Theta^{2}   \\
\mathcal{F}_{\eta}(z) &=\frac{1}{1-\frac{\eta^2 \alpha^2}{z^2}}.\label{E21}
\end{split}
\end{align}
Before we proceed further, a few important remarks regarding the $ \eta $ deformed background (\ref{E21}) are in order. One should notice that unlike the case for the usual $ AdS_2 $ background, the $ \eta $ deformed background (\ref{E21}) exhibits a metric singularity (associated with the function $ \mathcal{F}_{\eta}(z) $) at a finite radial distance,
\begin{eqnarray}
|z_B| =\eta~\alpha \equiv \frac{\eta}{J}\label{eee3.11}
\end{eqnarray}
which thereby naturally constraints our bulk calculations within this (radial) cutoff. This turns out to be a very special feature for spacetimes associated with YB deformations where one imagines putting the so called singularity surface namely the \textit{holographic screen} \cite{Kameyama:2014vma} at a finite radial distance $ z=z_B $ that eventually acts a boundary for the bulk spacetime. The boundary theory is therefore considered to be living on this holographic screen. Following the prescription of Gauge/Gravity duality, one could thereby imagine the above entity (\ref{eee3.11}) as being the energy scale associated with the holographic RG flow. Depending on the (radial) location of the holographic screen one is supposed to probe the physics associated with the dual field theory under the RG flow. The UV fixed point of this RG flow is given by the condition, $| \frac{\eta}{J}| \sim 0$ for which one recovers the metric corresponding to $ AdS_2 \times S^1 /Z_2 $. We now move on to the other end point of the RG flow where we set, $ |\frac{\eta}{J}|=\Lambda_{IR}\gg 1 $ where $ \Lambda_{IR} $ could be thought of as being that of the deep IR cutoff. In this limit, a careful analysis reveals the 2D bulk metric of the following form\footnote{For the moment, we ignore the (compact) third direction as it becomes trivial near both the fixed points.},
\begin{eqnarray}
d\mathsf{S}^{2}_{\eta}\Big|_{2D}\approx \frac{1}{|\varepsilon|^{2\varsigma}}(-d T^2 + d\varepsilon^{2})
\end{eqnarray}
where, we have introduced the following change of variables,
\begin{eqnarray}
\varepsilon =1-\frac{|\Lambda_{IR}|}{z},~~ T=\frac{t}{|\Lambda_{IR}|},~~|\varepsilon | \ll 1
\end{eqnarray}
together with the fact that the value corresponding to the dynamical critical exponent, $ \varsigma =\frac{1}{2} $. Therefore, in summary, the theory flows from a $ \varsigma =1 $ UV conformal fixed point to a $ \varsigma=\frac{1}{2} $ \textit{Lifshitz} fixed point in the deep IR.\\
In the following we (Fourier) decompose the scalar field as,
\begin{align}
\label{scalar.Four}
\phi = \int \frac{dw}{2 \pi}e^{-i w t}\xi_{w}(z,\Theta)
\end{align}
which finally yields the scalar action of the following form,
\begin{align}
\label{action.Four}
S_{\phi}=\frac{1}{2}\int dz d \Theta \int \frac{dw}{2\pi}\xi_{-w}\left( \mathfrak{D}_{0}
+\mathfrak{D}^{(\eta)} \right)\xi_{w}
\end{align}
where, the individual operators could be formally expressed as\footnote{Keeping the spirit of the earlier analysis \cite{Das:2017pif}, we have ignored all the higher order contributions 
beyond $ \mathcal{O}(\alpha^2 \eta^2) $. },
\begin{align}
\label{KK.opts}
\mathfrak{D}_{0}&=\partial_{z}^{2}+\omega^{2}-\frac{m^{2}}{z^{2}}-
\frac{1}{z^{2}}\left(-\partial_{\Theta}^{2}+V(\Theta) \right) \\
\mathfrak{D}^{(\eta)}&=\frac{\alpha}{z}\left[ \partial_{z}^{2}+\omega^{2}
-\frac{1}{z}\partial_{z} - \frac{m^{2}}{z^{2}}\left( 1+\frac{\alpha \eta^{2}}{z} \right)\right]
+\sum_{n=1,2} \left(-1\right)^{n}\frac{\alpha^{n}}{z^{n+2}} \mathsf{F}_{n}. 
\end{align}
Notice that like in the undeformed scenario \cite{Das:2017pif}, the zeroth order operator $ \mathfrak{D}_{0} $ (and hence the associated Green's function) does not receive any $ \eta $ corrections. However, the operator at next to leading order modifies substantially due to YB deformations,
\begin{eqnarray}
\mathsf{F}_{n}= \left(1+ \eta^{2(n-1)} \right) 
\partial_{\Theta}^{2}+(-1)^{n-1}\eta^{2(n-1)}V(\Theta). \label{E3.15}
\end{eqnarray}
For $ n=1 $ one has the usual contribution as in the undeformed case. Quite interestingly, the $ \eta $ modification appears with, $ n=2 $ which is an effect associated to $ \mathcal{O}(\alpha^2) $. Therefore, strictly at linear order in $ \alpha (\sim 1/J )$ \cite{Das:2017pif} one should not expect to find any of the imprints of YB deformations on the SYK spectrum.
\subsubsection{Eigenfunctions of $ \mathfrak{D}_{0} $}
The eigenfunctions corresponding to the operator $ \mathfrak{D}_{0} $ are clearly separable,
\begin{align}
\label{eigen.fun}
\xi_{w}(z,\Theta)=\mathcal{Z}_{w}(z)\mathfrak{f}_{\mathfrak{K}}(\Theta)
\end{align}
where the function $  \mathfrak{f}_{\mathfrak{K}}(\Theta)$ satisfies Schrodinger equation of the following form,
\begin{align}
(-\partial^{2}_{\Theta}+V \delta(\Theta))\mathfrak{f}_{\mathfrak{K}}(\Theta)=\mathsf{E}\mathfrak{f}_{\mathfrak{K}}(\Theta).\label{E27}
\end{align}
In our calculations, we stick to the \emph{parity even sector} \cite{Das:2017pif} of the wave function,
\begin{equation}
 \mathfrak{f}_{\mathfrak{K}}(\Theta)\sim
 \begin{cases}
\mathsf{B}  \sin (\mathfrak{K}(\Theta -L)), &~(\ 0<\Theta<L) \\
 -\mathsf{B} \sin (\mathfrak{K}(\Theta +L)), & ~ (-L<\Theta<0)
 \end{cases}\label{E28}
  \end{equation}
with, $ \mathsf{E}=\mathfrak{K}^2 $. Clearly, with the above choice of the wave function (\ref{E28}) one ends up with the following boundary conditions\footnote{For the moment we ignore the overall normalization constant.},
\begin{align}
\begin{split}
\mathfrak{f}_{\mathfrak{K}}(L)&=0=\mathfrak{f}_{\mathfrak{K}}(-L)    \\
\mathfrak{f}_{\mathfrak{K}}(+0)&=-\sin (\mathfrak{K}L)=\mathfrak{f}_{\mathfrak{K}}(-0).\label{E29}
\end{split}
\end{align}
Next, integrating (\ref{E27}) within a small interval around $ \Theta \sim 0 $ we notice,
\begin{align}
-\partial_{\Theta}\mathfrak{f}_{\mathfrak{K}}|_{-0}^{+0}+V \mathfrak{f}_{\mathfrak{K}}(0)
=\mathsf{E}(\mathfrak{f}_{\mathfrak{K}}(+0)-\mathfrak{f}_{\mathfrak{K}}(-0))
\end{align}
which upon substitution of (\ref{E29}) yields the following transcendental equation,
\begin{align}
-\frac{2}{V}\mathfrak{K}=\tan (\mathfrak{K}L)\label{E3.21}
\end{align}
whose solutions could be formally denoted as $ \mathsf{p}_a $  with $ a (=0,1,2,..)$ being an
integer and $ 2a+1<\mathsf{p}_a <2a+2 $ \cite{Das:2017pif}. 
\subsubsection{Eigenfunctions of $ \mathfrak{D}^{(\eta)} $}
Likewise, we separate the eigenfunctions of the operator $\mathfrak{D}^{(\eta)}$ as, 
\begin{equation}
\label{eigen.Dn}
\chi_{\omega}(z,\Theta)=\chi_{\omega}(z)\mathfrak{f}_{\tilde{\mathfrak{K}}_{n}}(\Theta)
\end{equation}
where, $\mathfrak{f}_{\tilde{\mathfrak{K}}_{n}}$ is the eigenfunction corresponding to the operator $ \mathsf{F}_{n} $ defined in (\ref{E3.15}),
\begin{eqnarray}
-\mathsf{F}_1 \mathfrak{f}_{\tilde{\mathfrak{K}}_{1}}(\Theta) = \tilde{\mathsf{E}}_{1}\mathfrak{f}_{\tilde{\mathfrak{K}}_{1}}(\Theta)\nonumber\\
\mathsf{F}_2 \mathfrak{f}_{\tilde{\mathfrak{K}}_{2}}(\Theta) = \tilde{\mathsf{E}}_{2}\mathfrak{f}_{\tilde{\mathfrak{K}}_{2}}(\Theta).
\end{eqnarray}
Restricting ourselves to the parity even sector of the wave function (\ref{E28}) it is trivial to arrive at the following set of transcendental equations,
\begin{eqnarray}
\frac{2}{V}\tilde{\mathfrak{K}}_{1}&=&\tan(\tilde{\mathfrak{K}}_{1}L)\label{E3.24}\\
-\frac{2}{V}\tilde{\mathfrak{K}}_{2}&=&\frac{\eta^{2}}{1+\eta^{2}}\tan(\tilde{\mathfrak{K}}_{2}L).\label{E3.25}
\end{eqnarray}
 \begin{figure}[h!]
    \centering
    {{\includegraphics[width=10cm]{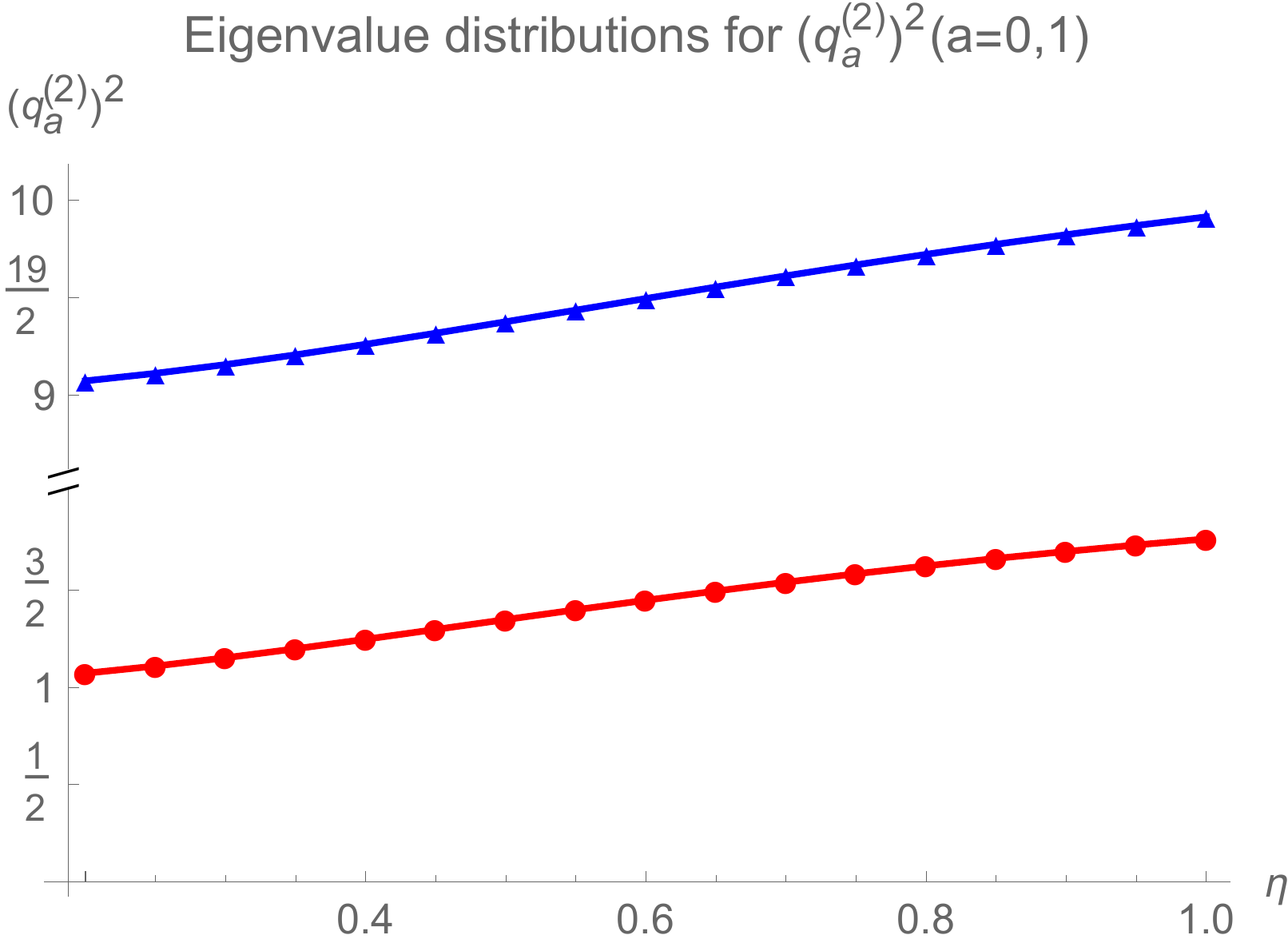} }}%
\caption{Energy eigenvalue spectrum with $ V=3 $ and $ L=\frac{\pi}{2} $. The blue curve corresponds to eigenvalues $ ( \mathsf{q}^{(2)}_{1}) ^2 $ and the red curve corresponds to eigenvalues $ ( \mathsf{q}^{(2)}_{0}) ^2 $. }
\label{Figure1}
\end{figure}
Notice that the second Eq.(\ref{E3.25}) makes sense \textit{only} in the presence of YB deformations ($ \eta \neq 0 $). As we shall see, the eigenfunctions corresponding to $ \mathsf{F}_{2} $ introduce a non trivial shift in the SYK spectrum at strong couplings. We denote the solutions corresponding to the above set of equations (\ref{E3.24})-(\ref{E3.25}) as $ 2a+1<\mathsf{q}^{(n)}_{a}<2a+2 $ where, $ n =1,2 $ refers to $ \mathsf{F}_{1} $ and $ \mathsf{F}_{2} $ respectively. We focus our attention on the first two roots ($ 1<\mathsf{q}^{(2)}_{0}<2  $ and $ 3<\mathsf{q}^{(2)}_{1}<4  $) corresponding to $ \mathsf{F}_{2} $ and explore their functional dependence with the deformation parameter ($ \eta $). To see this explicitly we plot the energy eigenvalues $ \tilde{\mathsf{E}}^{(a)}_{2}(\sim (\mathsf{q}^{(2)}_{a})^2 $) against the YB parameter ($ \eta $) (see, Fig.(\ref{Figure1})) which yields the following functional dependencies,
\begin{eqnarray}
\tilde{\mathsf{E}}^{(0)}_{2} \sim ( \mathsf{q}^{(2)}_{0}) ^2\approx  \mathcal{N}+\eta^{4/5},~~\mathcal{N}=0.778626
\end{eqnarray}
\begin{eqnarray}
\tilde{\mathsf{E}}^{(1)}_{2} \sim ( \mathsf{q}^{(2)}_{1}) ^2\approx \sum_{m=0}^{m=4}(-1)^m\mathcal{C}_{m}\eta^{m},~~\mathcal{C}_{m}\in \mathbb{R}.
\end{eqnarray}
Finally, substituting these eigenvalues into (\ref{KK.opts}) we obtain,
\begin{align}
\mathfrak{D}_{0}& = \partial^{2}_{z}+ w^2 -\frac{(m^{2}+\mathsf{p}^2_{a})}{z^2} ,  \label{D0} \\
\mathfrak{D}^{(\eta)}& =\frac{\alpha }{z} \left[\partial^2_z + w^{2} -\frac{1}{z} \partial_z 
-\sum_{n=1,2}\frac{\alpha^{n-1}}{z^{n+1}}\left( m^{2}\eta^{2(n-1)}-\mathsf{q}^{2(n)}_{a}\right) \right]. \label{Dn}
\end{align}
\subsection{Green's functions}
Given the desired choices for the parameters ($ V,L $) of the theory, we now compute the Green's function corresponding to the operator, $ \mathfrak{D}\equiv \mathfrak{D}_{0}+\mathfrak{D}^{(\eta)} $. It is naturally expected that the effects of YB deformations on the SYK spectrum would appear through the Green's function corresponding to the operator $ \mathfrak{D}^{(\eta)} $.
\subsubsection{Zeroth order solution}
The solution corresponding to the zeroth order propagator,
\begin{eqnarray}
\mathfrak{D}_{0}\mathcal{G}^{(0)}_{w,w'}(z,z';\Theta , \Theta')=-\delta (z-z')\delta (\Theta - \Theta')\delta (w+w')
\end{eqnarray}
remains the same as in the undeformed scenario\footnote{See Appendix A for details.}\cite{Das:2017pif}. Setting, $ \Theta =0=\Theta' $  into (\ref{A11}) one finally obtains,
\begin{eqnarray}
\mathcal{G}^{(0)}(t,z ;t',z')=-|zz'|^{1/2}\sum_{a=0}^{\infty}|\mathfrak{f}_{\mathsf{p}_{a}}(0)|^2\int \frac{dw}{2 \pi}e^{-iw(t-t')}\mathcal{I}(z,z')
\end{eqnarray}
where we note,
\begin{eqnarray}
|\mathfrak{f}_{\mathsf{p}_{a}}(0)|^2 &=& \mathsf{B}_{a}^{2}\sin^2(\mathfrak{K}L)=\frac{ \mathsf{B}_{a}^{2}\mathsf{p}^{2}_{a}}{(3/2)^2 +\mathsf{p}^2_{a}}=\frac{2 \mathsf{p}_a}{3}\mathsf{R}(\mathsf{p}_{a})\nonumber\\
\mathsf{R}(\mathsf{p}_{a})&=&\frac{3 \mathsf{p}^{2}_{a}}{((3/2)^2 +\mathsf{p}^2_{a})(\pi \mathsf{p}_{a}-\sin(\pi \mathsf{p}_{a}))}.
\end{eqnarray}
Setting, $ m^2 =-1/4 $ the integral $ \mathcal{I}(z,z')$ could be expressed as a sum of two terms\cite{Das:2017pif},
\begin{eqnarray}
\mathcal{I}_{1}=\sum_{n=0}^{\infty}\frac{2 \nu}{\nu^2 -\mathsf{p}^{2}_{a}}J_{\nu}(|wz|)J_{\nu}(|wz'|)\big|_{\nu=3/2+2n}\label{E3.33}
\end{eqnarray}
where, the index $ \nu $ is real. The above entity (\ref{E3.33}) has discrete poles at $ \nu =\mathsf{p}_{a} $ (on real $ \nu $ axis) and is clearly valid for $ \nu = 3/2+2n ~(n=0,1,2...) $ for which $ \xi_{\nu}=0 $ \cite{Polchinski:2016xgd}. This corresponds to bound energy (eigen)states with eigenvalues precisely as the roots of the transcendental equation (\ref{E3.21}). The second contribution to the integral $ \mathcal{I}(z,z')$ comes from the scattering states \cite{Polchinski:2016xgd} with $ \nu =ir~ (0<r<\infty) $ that amount of setting\footnote{The resulting wave function is a linear superposition of these two states \cite{Polchinski:2016xgd}.},
\begin{eqnarray}
\mathcal{I}_{2}=-\frac{i}{2}\int_{0}^{i \infty}\frac{d \nu}{\sin (\pi \nu)}\frac{\nu}{\nu^2 - \mathsf{p}^{2}_{a}}(\mathcal{Q}_{\nu}(z,z')+\tilde{\mathcal{Q}}_{\nu}(z,z'))\label{E3.34}
\end{eqnarray}
where, the individual entities in the integrand (\ref{E3.34}) could be formally expressed as,
\begin{eqnarray}
\mathcal{Q}_{\nu}(z,z')=(J_{-\nu}(|wz|)+\xi_{-\nu}J_{\nu}(|wz|))J_{\nu}(|wz'|)\nonumber\\
\tilde{\mathcal{Q}}_{\nu}(z,z')=(J_{\nu}(|wz|)+\xi_{\nu}J_{-\nu}(|wz|))J_{-\nu}(|wz'|).
\end{eqnarray}
A trivial change in the variable, $ \nu \rightarrow -\nu $ yields,
\begin{eqnarray}
\frac{i}{2}\int_{0}^{-i \infty}\frac{d \nu}{\sin (\pi \nu)}\frac{\nu}{\nu^2 - \mathsf{p}^{2}_{a}}\tilde{\mathcal{Q}}_{-\nu}(z,z')=-\frac{i}{2}\int_{-i \infty}^{0}\frac{d \nu}{\sin (\pi \nu)}\frac{\nu}{\nu^2 - \mathsf{p}^{2}_{a}}\mathcal{Q}_{\nu}(z,z')
\end{eqnarray} 
which thereby simplifies (\ref{E3.34}) as,
\begin{eqnarray}
\mathcal{I}_{2}=-\frac{i}{2}\int_{-i \infty}^{i \infty}\frac{d \nu}{\sin (\pi \nu)}\frac{\nu}{\nu^2 - \mathsf{p}^{2}_{a}}\mathcal{Q}_{\nu}(z,z';z>z').
\end{eqnarray}
We close the contour on the $ Re(\nu)>0 $ plane which essentially extends along the positive real axis of the complex $ \nu $ plane and extends to infinity along the imaginary axis. Clearly there are two types of poles within this contour- (I) simple pole at $ \nu=\mathsf{p}_{a} $ and (II) the poles (distributed along the real axis) associated with $ \xi_{-\nu} $ at $ \nu=3/2+2n $, 
\begin{eqnarray}
\mathcal{I}_{2}=-\frac{\pi}{2\sin(\pi \mathsf{p}_{a})}\mathcal{Q}_{\nu=\mathsf{p}_{a}}(z,z';z>z')-\mathcal{I}_{1}.\label{E3.38}
\end{eqnarray}
Combining (\ref{E3.33}) and (\ref{E3.38}) finally yields,
\begin{eqnarray}
\mathcal{G}^{(0)}(t,z ;t',z')=\frac{|zz'|^{1/2}}{4}\sum_{a=0}^{\infty}\frac{|\mathfrak{f}_{\mathsf{p}_{a}}(0)|^2}{\sin(\pi \mathsf{p}_{a})}\int dw e^{-iw(t-t')}\mathcal{Q}_{\nu=\mathsf{p}_{a}}(z,z';z>z').\label{E3.39}
\end{eqnarray}
\subsubsection{YB shift in the spectrum}
The purpose of this Section is to compute the first order shift in the energy spectrum corresponding to $ \mathfrak{D}^{(\eta)} $ and explore the effects of YB deformations at next to leading order in the SYK coupling ($ \alpha \sim  1/J$). A straightforward calculation reveals,
\begin{eqnarray}
\mathfrak{D}^{(\eta)}\sqrt{z}\mathcal{Z}_{\nu}(|wz|)=\frac{\alpha}{\sqrt{z}}\left( -\frac{\partial_{z}}{z}-\sum_{n=1,2}\frac{\alpha^{n-1}}{z^{n+1}}\Delta^{(n)}\right) \mathcal{Z}_{\nu}(|wz|).\label{E3.40}
\end{eqnarray}
where, we have introduced,
\begin{eqnarray}
\Delta^{(n)}=\left(\frac{3}{4}-\nu^2 \right) \delta_{n,1}+m^2 \eta^{2(n-1)}-\mathsf{q}^{2(n)}_{a}.
\end{eqnarray}
In order to simplify (\ref{E3.40}), we further notice that,
\begin{eqnarray}
\partial_{z}\mathcal{Z}_{\nu}(|wz|)=\partial_{z}J_{\nu}(|w z|)+\xi_{\nu}\partial_{z}J_{-\nu}(|w z|)
\end{eqnarray}
which by means of the following two identities,
\begin{eqnarray}
\partial_z J_{\nu}(z)&=&\frac{\nu}{z}J_{\nu}(z)-J_{\nu +1}(z)\nonumber\\
\partial_z J_{\nu}(z)&=&-\frac{\nu}{z}J_{\nu}(z)+J_{\nu -1}(z)\label{E3.42}
\end{eqnarray}
could be further simplified as\footnote{One has to replace, $ \nu \rightarrow -\nu $ in the second of the identities in (\ref{E3.42}).},
\begin{eqnarray}
\partial_{z}\mathcal{Z}_{\nu}(|wz|)=\frac{\nu}{|z|}\mathcal{Z}_{\nu}(|w z|)-|w| (J_{\nu+1}(|wz|)-\xi_{\nu}J_{-\nu -1}(|wz|)).\label{E3.43}
\end{eqnarray}
Substituting (\ref{E3.43}) into (\ref{E3.40}) we find,
\begin{eqnarray}
\mathfrak{D}^{(\eta)}\sqrt{z}\mathcal{Z}_{\nu}(|wz|)=-\frac{\alpha}{\sqrt{z}}\left( \frac{\nu}{z^2}+\sum_{n=1,2}\frac{\alpha^{n-1}}{z^{n+1}}\Delta^{(n)}\right) \mathcal{Z}_{\nu}(|wz|)\nonumber\\+\frac{\alpha}{z^{3/2}}|w|(J_{\nu+1}(|wz|)-\xi_{\nu}J_{-\nu -1}(|wz|))
\end{eqnarray}
which finally yields the matrix element of the following form,
\begin{eqnarray}
\int_{0}^{\infty}dz\sqrt{z}\mathcal{Z}_{\nu'}^{\ast}(|wz|)\mathfrak{D}^{(\eta)}\sqrt{z}\mathcal{Z}_{\nu}(|wz|)=<\mathsf{D}^{(\eta)}_{1}>+<\mathsf{D}^{(\eta)}_{2}>
\label{E3.40}
\end{eqnarray}
where the individual matrix elements could be formally expressed as,
\begin{eqnarray}
<\mathsf{D}^{(\eta)}_{1}>=-\alpha (\nu+\Delta^{(1)})\int_{0}^{\infty}\frac{dz}{z^2}\mathcal{Z}_{\nu'}^{\ast}(|wz|)\mathcal{Z}_{\nu}(|wz|)\nonumber\\
-\alpha^{2}\Delta^{(2)}\int_{0}^{\infty}\frac{dz}{z^3}\mathcal{Z}_{\nu'}^{\ast}(|wz|)\mathcal{Z}_{\nu}(|wz|)
\end{eqnarray}
\begin{eqnarray}
<\mathsf{D}^{(\eta)}_{2}>=\alpha |w|\int_{0}^{\infty}\frac{dz}{z}\mathcal{Z}_{\nu'}^{\ast}(|wz|)(J_{\nu+1}(|wz|)-\xi_{\nu}J_{-\nu -1}(|wz|)).
\end{eqnarray}
In order to evaluate the above integrals (\ref{E3.40}) we focus on the bound states with discrete energy eigenvalues corresponding to $ \nu =3/2+2n $ \cite{Polchinski:2016xgd}. This yields the following\footnote{We have set, $ m^2 =-1/4 $ which corresponds to setting the mass of the KK scalar at its BF bound. Also we have set, $ \nu' =\nu =3/2 $. This corresponds to the zero modes of the spectrum.},
\begin{eqnarray}
<\mathsf{D}^{(\eta)}>=\int_{0}^{\infty}dz\sqrt{z}\mathcal{Z}_{\nu'}^{\ast}(|wz|)\mathfrak{D}^{(\eta)}\sqrt{z}\mathcal{Z}_{\nu}(|wz|)=\frac{\alpha |w|}{2 \pi} (5/4+(\mathsf{q}^{(1)}_{0})^{2})+\Delta k_{YB}.\nonumber\\\label{E3.49}
\end{eqnarray}
The second term on the R.H.S. of (\ref{E3.49}) could be repackaged as,
\begin{eqnarray}
\Delta k_{YB} = \frac{\alpha^2 w^{2}}{30}(\eta^2 +4(\mathcal{N}+\eta^{4/5})^{2})\sim 1/J^2
\end{eqnarray}
is precisely the YB contribution to the SYK spectrum that appears as a next to leading order effect in the SYK coupling ($ \sim 1/J $). Notice that, here $ \Delta  k_{YB} (\geq 0)$ is manifestly positive definite (as a leading order effect in YB deformations) that shifts the pole of the propagator corresponding to the zero mode by an amount,
\begin{eqnarray}
\nu^2 =\left( \frac{3}{2}\right)^{2}+ <\mathsf{D}^{(\eta)}>
\end{eqnarray}
which could be further expanded as a perturbation in the SYK coupling,
\begin{eqnarray}
\nu = \frac{3}{2}+ \frac{|w|}{6\pi J}\Delta \nu_{(1)}+\frac{w^2}{90 J^2}\Delta \nu_{(2)}+\mathcal{O}(1/J^3).
\end{eqnarray}
Here, we have introduced new variables as,
\begin{eqnarray}
\Delta \nu_{(1)}&=&5/4+(\mathsf{q}^{(1)}_{0})^{2}\nonumber\\
\Delta \nu_{(2)}&=&\frac{30 J^2}{w^2}\Delta k_{YB}-\frac{5}{6 \pi^2}\Delta \nu_{(1)}^{2}.
\end{eqnarray} 
Notice that the shift $ \Delta \nu_{(2)} $ is purely a next to leading order ($ \sim 1/J^2 $) contribution to the spectrum that comes into play in the presence of YB deformations which forces us to consider effects beyond linear order in the perturbation expansion.
This finally yields the zero mode contribution to the propagator (\ref{E3.39}) as,
\begin{eqnarray}
\Delta \mathcal{G}^{(0)}(t,z ;t',z')&=&-\frac{9 \pi J \mathsf{B}^{2}_{0}}{4 }\frac{|zz'|^{1/2}}{\Delta \nu_{(1)}}\int_{-\infty}^{\infty} \frac{dw}{|w|}\Omega ^{-1}(J,w, \eta) e^{-i|w|(t-t')}J_{3/2}(|wz|)J_{3/2}(|w z'|)\nonumber\\
\Omega (J,w, \eta)&=&1+\frac{\pi w}{15 J}\frac{\Delta \nu_{(2)}}{•\Delta \nu_{(1)}}.
\end{eqnarray}
Clearly, if one switches off the YB deformation and restricts upto leading order ($ \sim 1/J $) in the perturbation series then, $\Omega =1  $ which reproduces the known results of \cite{Maldacena:2016hyu}.
\section{Bi-local holography and YB deformations}
From the analysis in the previous section, it is indeed quite evident that there are  precise \textit{holographic} confirmations of the perturbative energy shift in the SYK model due to the presence of YB deformations in the bulk counterpart. The purpose of this section is therefore to understand this deformation and in particular its consequences on the ($ 1+0 $)D SYK model in terms of bi-local excitations \cite{Jevicki:2016bwu} of the theory.
\subsection{Collective field excitations}
\subsubsection{Brief review of SYK}
The SYK model consists of $N(\gg 1)$ Majorana fermions with all-to-all interactions. This is a quantum mechanical model where the interactions between the fermions are completely random, described by a random coupling constant
$J_{ijkl}$ which exhibits a Gaussian distribution with zero mean ($  \langle J_{ijkl} \rangle =0$) and a non zero variance $\langle J_{ijkl}
J_{ijkl} \rangle = \frac{6 J^{2}}{N^{3}}$ \cite{Polchinski:2016xgd}. After performing the disorder averaging, there is only one coupling constant left in the theory \cite{Polchinski:2016xgd},
\begin{equation}
\label{dist.width}
J^{2}=\frac{N^{3}}{3!}\overline{J^{2}_{ijkl}}
\end{equation}
that appears in the effective action. 
The Hamiltonian of the system could be formally expressed as\footnote{The generalization of this model for $ q $ point vertex is quite straightforward \cite{Maldacena:2016hyu}.} \cite{Polchinski:2016xgd},
\begin{equation}
\label{hamiltonian.syk}
\mathcal{H}_{\text{SYK}}=\frac{1}{4!}\sum_{i,j,k,l=1}^{N}J_{ijkl}\chi_{i}\chi_{j}\chi_{k}\chi_{l}
\end{equation}
which leads to the Lagrangian of the following form,
\begin{equation}
\label{lagrangian.syk}
\mathcal{L}_{SYK}=  -\frac{1}{2!}\sum_{i=1}^{N}\chi_{i}\partial_{\tau}\chi_{i} - 
\frac{1}{4!}\sum_{i,j,k,l=1}^{N}J_{ijkl}\chi_{i}\chi_{j}\chi_{k}\chi_{l}.
\end{equation}
Here, $\chi_{i}$s ($i=1,2,\cdots,N$) are the so called Majorana fermions which satisfy the following anti-commutation
relation,
\begin{equation}
\label{anti.com}
\left\{ \chi_{i}, \, \chi_{j} \right\} = \delta_{ij}.
\end{equation}
together with the fact that they are equal to their own antiparticles namely, $\chi_{i}^{\dagger}=\chi_{i}$. 
One can usually calculate the free energy of the system by using the so called replica trick\footnote{Notice that the replica trick
is usually used in systems with quenched disorder in order to overcome the averaging over logarithms. Moreover, there is an amount of
ambiguity in taking the limit $n\rightarrow 0$ since we start from an integer $n$. Nevertheless, the 
results with replica method matches well with other methods.},
\begin{equation}
\label{repl.free-ener}
\langle  \ln \mathcal{Z} \rangle_{J} =  \lim_{n\rightarrow 0}\frac{\langle\mathcal{Z}^{n}\rangle_{J}-1}{n};
\qquad n\in \mathbb{Z}
\end{equation}
where, the partition function could be formally expressed as,
\begin{eqnarray}
\langle\mathcal{Z}^{n}\rangle_{J}=\int \mathcal{D}\chi_i ~ T( e^{\mathcal{S}^{(R)}}).
\end{eqnarray}
The action corresponding to the replica SYK is given by\footnote{The \textit{effective} replica action in (\ref{e4.7}) could be obtained after integrating over the random coupling,  $ J_{ijkl} $ in the path integral \cite{Jevicki:2016bwu}. The integral that one evaluates is Gaussian and is of the form, $ \sim \int \mathcal{D}\chi_i (..)\int \prod_{i,j,k,l=1}^{N} dJ_{ijkl}~e^{-N^3 J^2_{ijkl}/12 J^2} e^{\frac{1}{4!}J_{ijkl}\chi_i \chi_j \chi_k \chi_l}$.} \cite{Jevicki:2016bwu},
\begin{eqnarray}
\mathcal{S}^{(R)}=-\frac{1}{2}\sum_{a=1}^{n}\sum_{i=1}^{N}\int dt \chi^{a}_{i}(t)\partial_{t}\chi^{a}_{i}(t)-\frac{J^{2}}{8N^3}\int dt_1 dt_2 \sum_{a,b=1}^{n}\left(\sum_{i=1}^{N}\chi^{a}_{i}(t_1)\chi^{b}_{i}(t_2) \right)^{4}\label{e4.7}
\end{eqnarray}
where, the index $ a $ stands for the so called replica index.
In the large $ N $ limit, one introduces the \textit{bi-local} (collective) fields as \cite{Jevicki:2016bwu},
\begin{eqnarray}
\Psi (t_1 , t_2)=\frac{1}{N}\sum_{i=1}^{N}\chi_i (t_1)\chi_i (t_2)\label{e4.8}
\end{eqnarray}
where, we have suppressed the replica indices for later convenience. This yields the following path integral,
\begin{eqnarray}
\mathcal{Z}_{SYK}\sim \int \prod_{t_1 ,t_2}\mathcal{D}\Psi (t_1 ,t_2) \mu (\Psi)e^{-\mathcal{S}^{(R)}_{col}[\Psi]}
\end{eqnarray}
with the SYK replica action expressed in terms of collective field excitations (\ref{e4.8}),
\begin{eqnarray}
\mathcal{S}^{(R)}_{col}[\Psi]=\frac{N}{2}\int dt \partial_{t}\Psi (t,t')\Big|_{t'=t}+\frac{N}{2}Tr\log \Psi -\frac{J^2 N}{8}\int dt_1 dt_2 \Psi^{4}(t_1 , t_2).\label{e4.10}
\end{eqnarray}
Notice that since the collective field $ \Psi \sim \mathcal{O}(N^0) $, therefore the collective action in (\ref{e4.10}) is of $ \sim \mathcal{O}(N) $. The trace term in the action (\ref{e4.10}) takes care of the Jacobian that appears due to the change in the path integral measure from $ \chi_i (t) \rightarrow \Psi (t_1 , t_2) $. In the so called IR ($ |J t| \gg 1 $) limit one could ignore the kinematics of collective field excitations and therefore the collective action (\ref{e4.10}) is approximately reduced to,
\begin{eqnarray}
\mathcal{S}^{(IR)}_{col}[\Psi] \approx \frac{N}{2}Tr\log \Psi -\frac{J^2 N}{8}\int dt_1 dt_2 \Psi^{4}(t_1 , t_2)
\end{eqnarray}
which clearly has the re-parametrization invariance of the following form,
\begin{eqnarray}
t \rightarrow \tilde{t} = f(t);~~\Psi (t_1 ,t_2) \rightarrow (f'(t_1)f'(t_2))^{1/4}\tilde{\Psi}(\tilde{t}_{1},\tilde{t}_{2})
\end{eqnarray} 
characterizing the IR critical point at strong coupling.
\subsubsection{The effective action}
The classical saddle point equation has a solution \cite{Jevicki:2016bwu},
\begin{eqnarray}
\Psi_{0}(t_1 ,t_2)=-\left(\frac{1}{4 \pi J^2} \right)^{1/4} \frac{sgn(t_{12})}{\sqrt{|t_{12}|}};~~t_{12}=t_1 -t_2.
\end{eqnarray}
Next, we turn on \textit{fluctuations} around this IR fixed point and propose the following YB modification to the bi-local excitations,
\begin{eqnarray}
\Psi (t_1 ,t_2)=\Psi_{0}(t_1 ,t_2) +\sqrt{\frac{2}{N}}\zeta_{YB}
\end{eqnarray}
where the YB fluctuations associated to collective excitations\footnote{Since the YB deformation is expected to be appearing as $ 1/J^2 $ corrections to the SYK spectrum therefore it should not modify the solution ($ \Psi_{0}(t_1 ,t_2) $) corresponding to the IR critical point. },
\begin{eqnarray}
\zeta_{YB} (t_1 ,t_2)=\frac{\zeta (t_1 ,t_2)}{\mathcal{F}_{\eta}(t_1 ,t_2)};~~ \mathcal{F}_{\eta}(t_1 ,t_2)=\frac{1}{1-\frac{4 \eta^2}{J^2 (t_1 -t_2)^{2}}}
\end{eqnarray}
would precisely be identified with the corresponding scalar field d.o.f. living in the YB deformed $ AdS_2 $ spacetime. The collective action could be expanded around the IR critical point as,
\begin{eqnarray}
\mathcal{S}^{(R)}_{col}[\Psi]=\mathcal{S}^{(R)}_{col}[\Psi_{0}]+\frac{\delta^{2}\mathcal{S}^{(R)}_{col}}{\delta \zeta_{YB}(t_1 ,t_2)\delta \zeta_{YB}(t_3 ,t_4) }\delta \Psi (t_1 ,t_2) \delta \Psi (t_3 ,t_4)+\mathcal{O}(1/\sqrt{N})\nonumber\\\label{E4.16}
\end{eqnarray}
Clearly, the first term on the R.H.S. of (\ref{E4.16}) is of $ \mathcal{O}(N) $. Whereas, on the other hand, the second term is of $ \mathcal{O}(N^0) $ which could be formally expressed as,
\begin{eqnarray}
\mathcal{S}^{(2)}=\frac{1}{2}\int dt_1 dt_2 dt_3 dt_4~ \zeta_{YB}(t_1 ,t_2)\mathcal{K}(t_1,t_2;t_3,t_4)\zeta_{YB}(t_3 ,t_4)\label{e4.17}
\end{eqnarray}
where, the kernel could be formally expressed as \cite{Jevicki:2016bwu},
\begin{align}
\begin{split}
\mathcal{K}(X_{1},X_{2};X_{3},X_{4})&=-\frac{1}{2}\Big[\Psi^{-1}_{0}(X_{4},X_{1})\Psi^{-1}_{0}(X_{2},X_{3})
-\Psi^{-1}_{0}(X_{4},X_{2})\Psi^{-1}_{0}(X_{1},X_{3})\Big]    \\
&-3J^{2}\Psi_{0}(X_{1},X_{2})\Psi_{0}(X_{3},X_{4})\frac{1}{2}\Big(\delta_{X_{1},X_{3}}\delta_{X_{2},X_{4}}
+\delta_{X_{1},X_{4}}\delta_{X_{2},X_{3}} \Big).
\end{split}
\end{align}
In the following we evaluate the first term in\footnote{Here we have used the short hand notation, $ A \star B = \int dt A(t_a ,t)B(t ,t_b ) $ \cite{Jevicki:2016bwu}.} (\ref{e4.17}),
\begin{align}
\begin{split}
-\int dt_{1}dt_{2}&dt_{3}dt_{4}  \\
&\frac{1}{2}\zeta_{YB} (t_{1},t_{2})\Big[\Psi^{-1}_{0}(t_{4},t_{1})\Psi^{-1}_{0}(t_{2},t_{3})
-\Psi^{-1}_{0}(t_{4},t_{2})\Psi^{-1}_{0}(t_{1},t_{3})\Big]\zeta_{YB} (t_{3},t_{4})    \\
&=-\frac{1}{2}\int dt_{1}dt_{2}dt_{3}dt_{4}\Big[ \Psi^{-1}_{0}(t_{4},t_{1})\zeta_{YB} (t_{1},t_{2})\Psi^{-1}_{0}(t_{2},t_{3})\zeta_{YB} (t_{3},t_{4})  \\
&\qquad\qquad\qquad\qquad\qquad\qquad +\Psi^{-1}_{0}(t_{4},t_{2})\zeta_{YB} (t_{2},t_{1})\Psi^{-1}_{0}(t_{1},t_{3})\zeta_{YB} (t_{3},t_{4})\Big]   \\
&=-\int dt_{1}dt_{2}dt_{3}dt_{4}\;\Psi^{-1}_{0}(t_{4},t_{1})\zeta_{YB} (t_{1},t_{2})\Psi^{-1}_{0}(t_{2},t_{3})\zeta_{YB} (t_{3},t_{4})\\
&\equiv -\text{Tr}\left( \Psi_{0}^{-1}\star\zeta_{YB} \star \Psi_{0}^{-1}\star\zeta_{YB} \right) 
\end{split}\label{e4.19}
\end{align}
where, in the second line of (\ref{e4.19}) we have used the antisymmetry properties associated with the collective fields namely,
\begin{equation}
\label{antisym}
\Psi_{0}(X,Y)=-\Psi_{0}(Y,X),    \qquad\qquad\zeta_{YB}(X,Y)=-\zeta_{YB}(Y,X).
\end{equation}
This finally yields the effective quadratic action as,
\begin{eqnarray}
\mathcal{S}^{(2)}=-\frac{1}{2}\text{Tr}\left( \Psi_{0}^{-1}\star\zeta_{YB} \star \Psi_{0}^{-1}\star\zeta_{YB} \right) -
\frac{3J^{2}}{2}\int dt_{1}dt_{2}\left(\Psi_{0}(t_{1},t_{2})\right)^{2}\left(\zeta_{YB}(t_{1},t_{2})\right)^{2}.\label{e4.21}
\nonumber\\
\end{eqnarray}
In order to match our results to that with the $ (AdS_2)_{\eta} $ calculations in the preceeding section, we define the following coordinate transformations \cite{Jevicki:2016bwu},
\begin{eqnarray}
t=\frac{1}{2}(t_1 +t_2);~~z=\frac{1}{2}(t_1 -t_2)
\end{eqnarray}
and expand the fluctuations a complete basis,
\begin{eqnarray}
\psi (t,z)\equiv\zeta_{YB}(t_1 , t_2)=\sum_{\nu w}\tilde{\psi}_{\nu w}(t_1 ,t_2)b_{\nu w}(t_{1},t_{2});~~\tilde{\psi}_{\nu w}(t_1 ,t_2)=\frac{\psi_{\nu w}}{\mathcal{F}_{\eta}(t_1 ,t_2)}\label{e4.23}
\end{eqnarray}
that is defined as follows \cite{Jevicki:2016bwu},
\begin{eqnarray}
b_{\nu w}(t_{1},t_{2})=e^{iw\frac{t_{1}+t_{2}}{2}}\text{sgn}(t_{1}-t_{2})\mathcal{Z}_{\nu}\left(
\Big|w\frac{t_{1}-t_{2}}{2}\Big|\right)\label{e4.24}
\end{eqnarray}
where, $ \mathcal{Z}_{\nu}(|x|) $ is a linear combination of Bessel functions defined in (\ref{A9}). Following the original prescription of \cite{Jevicki:2016bwu},
\begin{eqnarray}
b_{\nu w}(t_1 ,t_2)=-\frac{3J}{16 \sqrt{\pi}}\frac{1}{g(\nu)}\int \frac{dt_a dt_b}{|t_a -t_b|}\Psi_0 (t_1 ,t_a)\Psi_0 (t_2 ,t_b)b_{\nu w}(t_a ,t_b)
\end{eqnarray}
one could further re-express (\ref{e4.23}) as\footnote{Here, $ \tilde{g}(\nu)=\frac{1}{g(\nu)}=-\frac{2 \nu}{3}\cot(\pi \nu /2) $\cite{Jevicki:2016bwu}.},
\begin{eqnarray}
\psi (t,z) = -\sum_{\nu w}\frac{3J \tilde{g}(\nu)}{16 \sqrt{\pi}}\tilde{\psi}_{\nu w}\int \frac{dt_a dt_b}{|t_a -t_b|}\Psi_0 (t_1 ,t_a) \Psi_0 (t_2 ,t_b)b_{\nu w}(t_a ,t_b).\label{e4.25}
\end{eqnarray}
Using (\ref{e4.25}), one could evaluate the quadratic action (\ref{e4.21}) as\footnote{See Appendix B for details of the analysis.},
\begin{eqnarray}
\mathcal{S}^{(2)}=\frac{3J}{32 \sqrt{\pi}}\sum_{\nu }\int dw~\vartheta_{\nu -w}N_{\nu} (\tilde{g}(\nu)-1)\vartheta_{\nu w}+\mathcal{O}(\frac{\eta^4}{J^4})\label{e4.27}
\end{eqnarray}
where, we have introduced,
\begin{eqnarray}
\vartheta_{\nu w}\approx \psi_{\nu w}\left( 1-\frac{\eta^2 w^2 }{8 \pi J^2 (\nu^2 -1)}\right).
\end{eqnarray}
The above Eq.(\ref{e4.27}) is one of the key findings of our paper. It turns out that the YB scaling associated with the bi-local fields in the SYK model produces a non trivial shift in the effective action (associated with quadratic fluctuations) at next to leading order ($ \sim 1/J^2 $) in the coupling which is thereby highly suppressed compared to that with the leading order ($ \sim 1/J $) effects. This observation is clearly compatible with our earlier findings in the previous section. This finally leads to the correlation function,
\begin{eqnarray}
\mathcal{G}(t_1,t_2;t_{1}',t_2')=\frac{16 \sqrt{\pi}}{3J}\sum_{\nu =3/2 +2n}\int dw \frac{b_{\nu -w}(t_1,t_2)b_{\nu w}(t_1',t_2')}{\tilde{N}_{\nu}(\tilde{g}(\nu)-1)}
\end{eqnarray}
where, we have defined,
\begin{eqnarray}
\tilde{N}_{\nu}=N_{\nu}\mathcal{F}_{\eta}(t_1 ,t_1')\mathcal{F}_{\eta}(t_2 ,t_2')\left( 1-\frac{\eta^2 w^2 }{4 \pi J^2 (\nu^2 -1)}\right).
\end{eqnarray}
\subsection{The $ (AdS_{2})_{\eta} $ spectrum}
Consider an effective action for scalar fields on $ AdS_2 $ \cite{Jevicki:2016bwu},
\begin{eqnarray}
S_{\varphi}&=&\frac{1}{2}\int d^2 x \sqrt{|g|}\left(-g^{ab}\partial_{a}\varphi_{m}\partial_{b}\varphi_{m}-\left(p^{2}_{m}-\frac{1}{4} \right)\varphi^{2}_{m}  \right)\nonumber\\
 &=&-\frac{1}{2}\int d^2 x \sqrt{|g|} \mathcal{L}_{\varphi}\label{E4.1}
\end{eqnarray}
where, the metric could be formally expressed as,
\begin{eqnarray}
g_{ab}&=&diag (-\mathcal{F}_{\eta}(z)/z^2 , \mathcal{F}_{\eta}(z)/z^2 )\nonumber\\
\mathcal{F}_{\eta}(z)&=&\frac{1}{1-\frac{\eta^2 \alpha^2}{z^2}}.
\end{eqnarray}
A straightforward calculation reveals,
\begin{eqnarray}
\sqrt{|g|} \mathcal{L}_{\phi}=-\varphi_{m}\left( \square^{(2)} -\frac{\mathcal{F}_{\eta}(z)}{z^2}\left( p^{2}_{m}-\frac{1}{4}\right)\right)\varphi_{m}.
\end{eqnarray}
On the other hand, it is trivial to notice that\footnote{Here, $ D_B = z^2 \partial^{2}_{z}+z\partial_{z}-z^2\partial^{2}_{t} $ is the Bessel differential operator.} \cite{Das:2017pif},
\begin{eqnarray}
\nabla^{a}\nabla_{a}\varphi_{m} = \frac{z^2}{\mathcal{F}_{\eta}(z)}\square^{(2)}\varphi_{m} =\sqrt{z}\mathrm{D}_{B}\left(\frac{\tilde{\varphi}_{m}}{\sqrt{z}} \right)-\frac{\tilde{\varphi}_{m}}{4} \tilde{\mathcal{F}}_{\eta}(z)
\end{eqnarray}
where, we have rescaled the scalar field as\footnote{This precisely confirms that the bi-local fields in the SYK should also get appropriately rescaled in the presence of YB deformations.}, $\tilde{\varphi}_{m} =\frac{\varphi_{m}}{\mathcal{F}_{\eta}(z)}$. Notice that here we have introduced a new function,
\begin{eqnarray}
\tilde{\mathcal{F}}_{\eta}(z)=\mathcal{F}_{\eta}(z)\left( 1-\frac{17 \alpha^2 \eta^2}{z^2}\right) 
\end{eqnarray}
that precisely goes to unity in the limit of the vanishing YB deoformations and thereby one recovers the original results of \cite{Das:2017pif}. Based on the above analysis, we propose the following \textit{non local} field redefinition,
\begin{eqnarray}
\varphi_{m} (t,z)= \left( \frac{3J}{8 \sqrt{\pi}}\right)^{1/2} z^{1/2}\sqrt{f(\sqrt{D_B})}~\tilde{\varphi}_{m}(t,z).\label{E4.6}
\end{eqnarray}
Substituting (\ref{E4.6}) into (\ref{E4.1}) we obtain,
\begin{eqnarray}
S_{\varphi}=\frac{1}{2}\frac{3J}{8 \sqrt{\pi}}\int dt \int_{0}^{\infty}\frac{dz}{z}\tilde{\varphi}_{m}(D_B -\tilde{p}^{2}_{m})f\tilde{\varphi}_{m}\label{E4.7}
\end{eqnarray}
where, the pole has been rescaled due to YB deformations as,
\begin{eqnarray}
\tilde{p}^{2}_{m}=\mathcal{F}_{\eta}(z)\left(p^2_{m}-\frac{\eta^2 \alpha^2}{4z^2} \right).\label{E4.8}
\end{eqnarray}
Implementing the definition \cite{Jevicki:2016bwu},
\begin{eqnarray}
\tilde{g}(\nu)-1=(\nu^2  -\tilde{p}^{2}_{m})f(\nu)
\end{eqnarray}
one could further express (\ref{E4.7}) as,
\begin{eqnarray}
S_{\varphi}&=&\frac{1}{2}\frac{3J}{8 \sqrt{\pi}}\int dt \int_{0}^{\infty}\frac{dz}{z}\tilde{\varphi}_{m}\left(\tilde{g}(\sqrt{D_B})-1 \right) \tilde{\varphi}_{m}\nonumber\\&=&\frac{1}{2}\frac{3J}{8 \sqrt{\pi}}\int dt \int_{0}^{\infty}\frac{dz}{z}\tilde{\Phi}_{m}\left(\tilde{g}(\sqrt{D_B})-1 \right) \tilde{\Phi}_{m}+\mathcal{O}(\eta^4 \alpha^4)\label{E4.10}
\end{eqnarray}
where, we have introduced,
\begin{eqnarray}
\tilde{\Phi}_{m} \approx \varphi_{m}\left( 1-\frac{\eta^2 \alpha^2}{z^2}\right).
\end{eqnarray}
Notice that the above equation (\ref{E4.10}) clearly resemblance our previous finding in (\ref{e4.27}). It is also worthwhile to mention that in the limit of the vanishing YB deformations, the quadratic action (\ref{E4.10}) precisely reproduces the previous findings of \cite{Jevicki:2016bwu}. It is indeed interesting to notice that, as observed in the previous section, the YB deformations shifts the pole  (\ref{E4.8}) by an amount that goes with the quadratic order ($ \sim 1/J^2 $) in the inverse of the SYK coupling. 
\section{Concluding remarks}
In this paper, based on the notion of SYK/AdS correspondence, we explore the effects of Yang-Baxter (YB) deformations on the collective field excitations within the SYK model. The motivation behind our analysis solely comes from the underlying holographic principle which strongly suggests a possible modification of the SYK spectrum at quadratic ($ 1/J^2 $) order in the SYK coupling. Based on the notion of holography (namely, looking at the scalar fluctuations and their YB scaling in $ (AdS_2)_{\eta} $) we propose a possible YB scaling of the bi-local fields in the SYK model and compute the effective action upto quadratic order in the fluctuations. It would be really nice to understand this YB scaling in terms of $ 1/N $ diagrammatics and thereby the associated Feynman rules in terms of these newly defined collective excitations. We hope to address some of these issues in the near future.   
\acknowledgments
It is indeed a great pleasure to thank Kenta Suzuki for valuable correspondences on several technical aspects of the manuscript. The authors would also like to convey their sincere thanks to Sumit R. Das, Antal Jevicki, Marika Taylor and Kenta Suzuki for their valuable comments on the draft. AL would like to acknowledge the financial support from PUCV, Chile. The work of
 DR was supported through the Newton-Bhahba Fund. DR would like to acknowledge \emph{the 
Royal Society} UK and \emph{the Science and Engineering Research Board India} (SERB) for financial assistance.
\appendix
\section{Evaluation of the Green's function $\mathcal{G}^{(0)}_{w,w'}$}
The zero-th order Green's function $\mathcal{G}^{(0)}_{w,w'}$ is defined through the following equation,
\begin{align}
\mathfrak{D}_{0}\mathcal{G}^{(0)}_{w,w'}(z,z';\Theta , \Theta')=-\delta (z-z')\delta (\Theta - \Theta')\delta (w+w'). \label{A1}
\end{align}
Expressing the Green's function in a basis of orthonormal wave functions, 
\begin{eqnarray}
\mathcal{G}^{(0)}_{w,w'}=\sqrt{z}\sum_{\mathfrak{K},\mathfrak{K}'}\mathfrak{f}_{\mathfrak{K}}(\Theta)
\mathfrak{f}_{\mathfrak{K}'}(\Theta')\tilde{G}^{(0)}_{\omega,\mathfrak{K};\omega',\mathfrak{K}'}(z;z')
\end{eqnarray}
and substituting back into (\ref{A1}) we find,
\begin{equation}
\left( \hat{\mathcal{L}}_{B}-
\nu_{0}^{2} \right)\tilde{G}^{(0)}_{w,\mathfrak{K};-w,\mathfrak{K}}(z;z')=-z^{3/2}\delta(z-z')
   \label{A3}
\end{equation}
where we have introduced,
\begin{eqnarray}
\hat{\mathcal{L}}_{B}& =& z^2 \partial^2_{z}+z\partial_z + w^2 z^2 \\
\nu_{0}^{2}&=&\mathsf{p}_{a}^{2}+m^{2}+\frac{1}{4}  
\end{eqnarray}
and used the orthonormality conditions for the wave functions,
\begin{eqnarray}
\sum_{\mathfrak{K},\mathfrak{K}'}\mathfrak{f}_{\mathfrak{K}}(\Theta)
\mathfrak{f}_{\mathfrak{K}'}(\Theta')=\delta_{\mathfrak{K},\mathfrak{K}'}\delta\left(\Theta-\Theta'\right).
\end{eqnarray}
We express the Green's function (\ref{A3}) in a basis of Bessel function,
\begin{eqnarray}
\tilde{G}^{(0)}_{w,\mathfrak{K};-w,\mathfrak{K}}(z;z') =\int d \nu \tilde{\mathfrak{g}}^{(0)}_{\nu}(z')\mathcal{Z}_{\nu}(| w z|)
\end{eqnarray}
that satisfies the Bessel equation,
\begin{eqnarray}
 \hat{\mathcal{L}}_{B}\mathcal{Z}_{\nu}(| w z|) = \nu^2 \mathcal{Z}_{\nu}(| w z|).\label{A8}
\end{eqnarray}
The most general solution to (\ref{A8}) could be formally expressed as \cite{Polchinski:2016xgd},
\begin{eqnarray}
\mathcal{Z}_{\nu}(|x|)=J_{\nu}(|x|)+\xi_{\nu}J_{-\nu}(|x|).\label{A9}
\end{eqnarray}
Notice that while both functions converge at large $ |x|\rightarrow \infty $, one of the solutions $ J_{-\nu}(x) $ diverges for $ x\sim 0 $ which thereby amounts of setting the coefficient,
\begin{eqnarray}
\xi_{\nu}=\frac{\tan(\pi \nu/2)+1}{\tan(\pi \nu/2)-1}=0\Rightarrow ~~\nu =3/2+2n.
\end{eqnarray} 
Substituting, (\ref{A8}) into (\ref{A3}) and using the completeness condition,
\begin{eqnarray}
\int \frac{d \nu}{N_{\nu}}\mathcal{Z}_{\nu}^{\ast}(|x|) \mathcal{Z}_{\nu}(|x'|) =x \delta (x-x')
\end{eqnarray}
it is in fact quite straightforward to show,
\begin{eqnarray}
\tilde{\mathfrak{g}}^{(0)}_{\nu}(z')=-\frac{\sqrt{z'}}{N_{\nu}}\mathcal{Z}_{\nu}^{\ast}(|wz'|)
\end{eqnarray}
which finally yields the real space zeroth order Green's function,
\begin{eqnarray}
\mathcal{G}^{(0)}(t,z,\Theta ;t',z',\Theta')&=&-|zz'|^{1/2}\sum_{a=0}^{\infty}\mathfrak{f}_{\mathsf{p}_{a}}(\Theta)\mathfrak{f}_{\mathsf{p}_{a}}(\Theta')\int \frac{dw}{2 \pi}e^{-iw(t-t')}\mathcal{I}(z,z')\nonumber\\
\mathcal{I}(z,z')&=&\int \frac{d \nu}{N_{\nu}}\frac{\mathcal{Z}_{\nu}^{\ast}(|wz|) \mathcal{Z}_{\nu}(|w z'|)}{\nu^2 -\nu_{0}^{2}}.\label{A11}
\end{eqnarray}
\section{Evaluation of the quadratic action $\mathcal{S}^{(2)}$}
We divide the quadratic action (\ref{e4.21}) into following two parts,
\begin{eqnarray}
\mathcal{S}^{(2)}_{I}=-\frac{1}{2}\int dt_{1}dt_{2}dt_{3}dt_{4}\;\Psi^{-1}_{0}(t_{4},t_{1})\zeta_{YB} (t_{1},t_{2})\Psi^{-1}_{0}(t_{2},t_{3})\zeta_{YB} (t_{3},t_{4})\nonumber\\
=-\frac{1}{2}\sum_{\nu ,\nu'}\int dw dw' dt_1 dt_2 dt_3 dt_4 \Psi_0^{-1}(t_4 ,t_1) \Psi_0^{-1}(t_2 ,t_3)~~~~~~~~~~~~~\nonumber\\
\tilde{\psi}_{\nu w}(t_1 ,t_2)\tilde{\psi}_{\nu' w'}(t_3 ,t_4)b_{\nu w}(t_1 ,t_2)b_{\nu' w'}(t_3, t_4)\nonumber\\
=\frac{3J}{32 \sqrt{\pi}}\sum_{\nu ,\nu'}\tilde{g}(\nu')\int dw dw' dt_1 dt_2 dt_3 dt_4\frac{dt_a dt_b}{|t_a -t_b|}\Psi^{-1}_{0}(t_{1},t_{4})\Psi_{0}(t_{4},t_{b})\nonumber\\
\Psi^{-1}_{0}(t_{2},t_{3})\Psi_{0}(t_{3},t_{a})\tilde{\psi}_{\nu w}(t_1 ,t_2)\tilde{\psi}_{\nu' w'}(t_3 ,t_4)b_{\nu w}(t_1 ,t_2)b_{\nu' w'}(t_b, t_a)\nonumber\\
=\frac{3J}{32 \sqrt{\pi}}\sum_{\nu ,\nu'}\tilde{g}(\nu')\int dw dw' \frac{dt_1 dt_2}{|t_1 -t_2|}\psi_{\nu w}\psi_{\nu' w'}b_{\nu w}(t_1 ,t_2)b_{\nu' w'}(t_1, t_2)\nonumber\\
-\frac{3 \eta^2}{4\sqrt{\pi}J}\sum_{\nu ,\nu'}\tilde{g}(\nu')\int dw dw' \frac{dt_1 dt_2}{|t_1 -t_2|^{3}}\psi_{\nu w}\psi_{\nu' w'}b_{\nu w}(t_1 ,t_2)b_{\nu' w'}(t_1, t_2)+\mathcal{O}(\frac{\eta^4}{J^3})
\end{eqnarray}
where, we have used the orthogonalization condition \cite{Polchinski:2016xgd},
\begin{eqnarray}
\int dt \Psi^{-1}_{0}(t_a ,t)\Psi_{0}(t, t_b)=\delta (t_a -t_b).
\end{eqnarray}
Using (\ref{e4.24}) this could be further re-expressed as,
\begin{eqnarray}
\mathcal{S}^{(2)}_{I}=\frac{3J}{32 \sqrt{\pi}}\sum_{\nu }\int dw  \psi_{\nu -w}N_{\nu}\left( 1-\frac{\eta^2 w^2 }{4 \pi J^2 (\nu^2 -1)}\right) \tilde{g}(\nu)\psi_{\nu w}+\mathcal{O}(\frac{\eta^4}{J^4})
\end{eqnarray}
where, we have performed the integral only for bound states with integer $ \nu = \frac{3}{2}+2n~(n=0,1,2..) $.
\begin{eqnarray}
\mathcal{S}^{(2)}_{II}=-
\frac{3J^{2}}{2}\frac{1}{8}\int dt_{1}dt_{2}\left(\Psi_{0}(t_{1},t_{2})\right)^{2}\left(\zeta_{YB}(t_{1},t_{2})\right)^{2}~~~~~~~~~~~~~~~\nonumber\\
=-\frac{3J^{2}}{2}\frac{1}{8}\sum_{\nu ,\nu'}\int dw dw' dt_1 dt_2 \left(\Psi_{0}(t_{1},t_{2})\right)^{2} \tilde{\psi}_{\nu w}(t_1 , t_2)~~~~\nonumber\\
\tilde{\psi}_{\nu' w'}(t_1 ,t_2)b_{\nu w}(t_1 ,t_2)b_{\nu' w'}(t_1 ,t_2)\nonumber\\
=-\frac{3J^{2}}{2}\frac{1}{8}\sum_{\nu ,\nu'}\int dw dw' dt_1 dt_2 \left(\Psi_{0}(t_{1},t_{2})\right)^{2} \psi_{\nu w}\psi_{\nu' w'}~~~~~\nonumber\\
b_{\nu w}(t_1 ,t_2)b_{\nu' w'}(t_1 ,t_2)\nonumber\\
+12\eta^{2}\frac{1}{8}\sum_{\nu ,\nu'}\int dw dw' \frac{dt_1 dt_2}{|t_1 -t_2|^2} \left(\Psi_{0}(t_{1},t_{2})\right)^{2} \psi_{\nu w}\psi_{\nu' w'}\nonumber\\
b_{\nu w}(t_1 ,t_2)b_{\nu' w'}(t_1 ,t_2)+\mathcal{O}(\frac{\eta^4}{J^3})\nonumber\\
=-\frac{3J}{32 \sqrt{\pi}}\sum_{\nu }\int dw\psi_{\nu -w}N_{\nu}\left( 1-\frac{\eta^2 w^2 }{4 \pi J^2 (\nu^2 -1)}\right) \psi_{\nu w}+\mathcal{O}(\frac{\eta^4}{J^4})
\end{eqnarray}
where, the factor $ 1/8 $ has been introduced in order to avoid the overcounting in the expansion of $ \sum_{i,j=1}^{4} \Psi_{0}(X_i ,X_j)^{2}\eta (X_i , X_j)^{2}$.


\begin{thebibliography}{99}
\bibitem{Sachdev:1992fk} 
  S.~Sachdev and J.~Ye,
  ``Gapless spin fluid ground state in a random, quantum Heisenberg magnet,''
  Phys.\ Rev.\ Lett.\  {\bf 70}, 3339 (1993)
  doi:10.1103/PhysRevLett.70.3339
  [cond-mat/9212030].

\bibitem{Sachdev:2010um} 
  S.~Sachdev,
  ``Holographic metals and the fractionalized Fermi liquid,''
  Phys.\ Rev.\ Lett.\  {\bf 105}, 151602 (2010)
  doi:10.1103/PhysRevLett.105.151602
  [arXiv:1006.3794 [hep-th]].
  
  \bibitem{Sachdev:2010uj} 
  S.~Sachdev,
  ``Strange metals and the AdS/CFT correspondence,''
  J.\ Stat.\ Mech.\  {\bf 1011}, P11022 (2010)
  doi:10.1088/1742-5468/2010/11/P11022
  [arXiv:1010.0682 [cond-mat.str-el]].
  
  \bibitem{Kitaev1}A. Kitaev. 2015. A simple model of quantum holography, talk given at KITP strings seminar and Entanglementprogram, February 12, April 7, and May 27, Santa Barbara, U.S.A.
  \bibitem{Kitaev2}A. Kitaev. 2014. Hidden correlations in the Hawking radiation and thermal noise, talk given at Fundamental Physics Prize Symposium, November 10, Santa Barbara, U.S.A.
  
  \bibitem{Sachdev:2015efa} 
  S.~Sachdev,
  ``Bekenstein-Hawking Entropy and Strange Metals,''
  Phys.\ Rev.\ X {\bf 5}, no. 4, 041025 (2015)
  doi:10.1103/PhysRevX.5.041025
  [arXiv:1506.05111 [hep-th]].

\bibitem{Polchinski:2016xgd} 
  J.~Polchinski and V.~Rosenhaus,
  ``The Spectrum in the Sachdev-Ye-Kitaev Model,''
  JHEP {\bf 1604}, 001 (2016)
  doi:10.1007/JHEP04(2016)001
  [arXiv:1601.06768 [hep-th]].
  
  \bibitem{Maldacena:2016hyu} 
  J.~Maldacena and D.~Stanford,
  ``Remarks on the Sachdev-Ye-Kitaev model,''
  Phys.\ Rev.\ D {\bf 94}, no. 10, 106002 (2016)
  doi:10.1103/PhysRevD.94.106002
  [arXiv:1604.07818 [hep-th]].
  
  \bibitem{Fu:2016vas} 
  W.~Fu, D.~Gaiotto, J.~Maldacena and S.~Sachdev,
  ``Supersymmetric Sachdev-Ye-Kitaev models,''
  Phys.\ Rev.\ D {\bf 95}, no. 2, 026009 (2017)
  Addendum: [Phys.\ Rev.\ D {\bf 95}, no. 6, 069904 (2017)]
  doi:10.1103/PhysRevD.95.069904, 10.1103/PhysRevD.95.026009
  [arXiv:1610.08917 [hep-th]].
  
  \bibitem{Yoon:2017gut} 
  J.~Yoon,
  ``Supersymmetric SYK Model: Bi-local Collective Superfield/Supermatrix Formulation,''
  JHEP {\bf 1710}, 172 (2017)
  doi:10.1007/JHEP10(2017)172
  [arXiv:1706.05914 [hep-th]].
  
  \bibitem{Garcia-Garcia:2016mno} 
  A.~M.~Garcia-Garcia and J.~J.~M.~Verbaarschot,
  ``Spectral and thermodynamic properties of the Sachdev-Ye-Kitaev model,''
  Phys.\ Rev.\ D {\bf 94}, no. 12, 126010 (2016)
  doi:10.1103/PhysRevD.94.126010
  [arXiv:1610.03816 [hep-th]].
  
  \bibitem{Garcia-Garcia:2017pzl} 
  A.~M.~Garcia-Garcia and J.~J.~M.~Verbaarschot,
  ``Analytical Spectral Density of the Sachdev-Ye-Kitaev Model at finite N,''
  Phys.\ Rev.\ D {\bf 96}, no. 6, 066012 (2017)
  doi:10.1103/PhysRevD.96.066012
  [arXiv:1701.06593 [hep-th]].
  
  \bibitem{Jevicki:2016bwu} 
  A.~Jevicki, K.~Suzuki and J.~Yoon,
  ``Bi-Local Holography in the SYK Model,''
  JHEP {\bf 1607}, 007 (2016)
  doi:10.1007/JHEP07(2016)007
  [arXiv:1603.06246 [hep-th]].
  
  \bibitem{Jevicki:2016ito} 
  A.~Jevicki and K.~Suzuki,
  ``Bi-Local Holography in the SYK Model: Perturbations,''
  JHEP {\bf 1611}, 046 (2016)
  doi:10.1007/JHEP11(2016)046
  [arXiv:1608.07567 [hep-th]].
  
  \bibitem{Gross:2016kjj} 
  D.~J.~Gross and V.~Rosenhaus,
  ``A Generalization of Sachdev-Ye-Kitaev,''
  JHEP {\bf 1702}, 093 (2017)
  doi:10.1007/JHEP02(2017)093
  [arXiv:1610.01569 [hep-th]].
  
  \bibitem{Gross:2017hcz} 
  D.~J.~Gross and V.~Rosenhaus,
  ``The Bulk Dual of SYK: Cubic Couplings,''
  JHEP {\bf 1705}, 092 (2017)
  doi:10.1007/JHEP05(2017)092
  [arXiv:1702.08016 [hep-th]].
  
  \bibitem{Kitaev:2017awl} 
  A.~Kitaev and S.~J.~Suh,
  ``The soft mode in the Sachdev-Ye-Kitaev model and its gravity dual,''
  JHEP {\bf 1805}, 183 (2018)
  doi:10.1007/JHEP05(2018)183
  [arXiv:1711.08467 [hep-th]].
  
  \bibitem{Krishnan:2016bvg} 
  C.~Krishnan, S.~Sanyal and P.~N.~Bala Subramanian,
  ``Quantum Chaos and Holographic Tensor Models,''
  JHEP {\bf 1703}, 056 (2017)
  doi:10.1007/JHEP03(2017)056
  [arXiv:1612.06330 [hep-th]];~
  M.~Berkooz, P.~Narayan, M.~Rozali and J.~Simon,
  ``Higher Dimensional Generalizations of the SYK Model,''
  JHEP {\bf 1701}, 138 (2017)
  doi:10.1007/JHEP01(2017)138
  [arXiv:1610.02422 [hep-th]].
  
  
  \bibitem{Peng:2017spg} 
  C.~Peng, M.~Spradlin and A.~Volovich,
  ``Correlators in the $\mathcal{N}=2$ Supersymmetric SYK Model,''
  JHEP {\bf 1710}, 202 (2017)
  doi:10.1007/JHEP10(2017)202
  [arXiv:1706.06078 [hep-th]].
  
  \bibitem{Peng:2016mxj} 
  C.~Peng, M.~Spradlin and A.~Volovich,
  ``A Supersymmetric SYK-like Tensor Model,''
  JHEP {\bf 1705}, 062 (2017)
  doi:10.1007/JHEP05(2017)062
  [arXiv:1612.03851 [hep-th]].
  
 \bibitem{Taylor:2017dly} 
  M.~Taylor,
  ``Generalized conformal structure, dilaton gravity and SYK,''
  JHEP {\bf 1801}, 010 (2018)
  doi:10.1007/JHEP01(2018)010
  [arXiv:1706.07812 [hep-th]].
  
  \bibitem{Forste:2017apw} 
  S.~Forste, J.~Kames-King and M.~Wiesner,
  ``Towards the Holographic Dual of N = 2 SYK,''
  JHEP {\bf 1803}, 028 (2018)
  doi:10.1007/JHEP03(2018)028
  [arXiv:1712.07398 [hep-th]].
  
  \bibitem{Rosenhaus:2018dtp} 
  V.~Rosenhaus,
  ``An introduction to the SYK model,''
  arXiv:1807.03334 [hep-th].
  
  \bibitem{Shenker:2013pqa} 
  S.~H.~Shenker and D.~Stanford,
  ``Black holes and the butterfly effect,''
  JHEP {\bf 1403}, 067 (2014)
  doi:10.1007/JHEP03(2014)067
  [arXiv:1306.0622 [hep-th]].
  
  \bibitem{Shenker:2014cwa} 
  S.~H.~Shenker and D.~Stanford,
  ``Stringy effects in scrambling,''
  JHEP {\bf 1505}, 132 (2015)
  doi:10.1007/JHEP05(2015)132
  [arXiv:1412.6087 [hep-th]].
  
  \bibitem{Maldacena:2015waa} 
  J.~Maldacena, S.~H.~Shenker and D.~Stanford,
  ``A bound on chaos,''
  JHEP {\bf 1608}, 106 (2016)
  doi:10.1007/JHEP08(2016)106
  [arXiv:1503.01409 [hep-th]].
  
  \bibitem{Teitelboim:1983ux} 
  C.~Teitelboim,
  ``Gravitation and Hamiltonian Structure in Two Space-Time Dimensions,''
  Phys.\ Lett.\  {\bf 126B}, 41 (1983).
  doi:10.1016/0370-2693(83)90012-6
  
  \bibitem{Jackiw:1984je} 
  R.~Jackiw,
  ``Lower Dimensional Gravity,''
  Nucl.\ Phys.\ B {\bf 252}, 343 (1985).
  doi:10.1016/0550-3213(85)90448-1
  
  \bibitem{Almheiri:2014cka} 
  A.~Almheiri and J.~Polchinski,
  ``Models of AdS$_{2}$ backreaction and holography,''
  JHEP {\bf 1511}, 014 (2015)
  doi:10.1007/JHEP11(2015)014
  [arXiv:1402.6334 [hep-th]].
  
  \bibitem{Maldacena:2016upp} 
  J.~Maldacena, D.~Stanford and Z.~Yang,
  ``Conformal symmetry and its breaking in two dimensional Nearly Anti-de-Sitter space,''
  PTEP {\bf 2016}, no. 12, 12C104 (2016)
  doi:10.1093/ptep/ptw124
  [arXiv:1606.01857 [hep-th]].
  
  \bibitem{Cvetic:2016eiv} 
  M.~Cvetic and I.~Papadimitriou,
  ``AdS$_{2}$ holographic dictionary,''
  JHEP {\bf 1612}, 008 (2016)
  Erratum: [JHEP {\bf 1701}, 120 (2017)]
  doi:10.1007/JHEP12(2016)008, 10.1007/JHEP01(2017)120
  [arXiv:1608.07018 [hep-th]].
  
  \bibitem{Mandal:2017thl} 
  G.~Mandal, P.~Nayak and S.~R.~Wadia,
  ``Coadjoint orbit action of Virasoro group and two-dimensional quantum gravity dual to SYK/tensor models,''
  JHEP {\bf 1711}, 046 (2017)
  doi:10.1007/JHEP11(2017)046
  [arXiv:1702.04266 [hep-th]].
  
  \bibitem{Engelsoy:2016xyb} 
  J.~Engelsoy, T.~G.~Mertens and H.~Verlinde,
  ``An investigation of AdS$_{2}$ backreaction and holography,''
  JHEP {\bf 1607}, 139 (2016)
  doi:10.1007/JHEP07(2016)139
  [arXiv:1606.03438 [hep-th]];~K.~Jensen,
  ``Chaos in AdS$_2$ Holography,''
  Phys.\ Rev.\ Lett.\  {\bf 117}, no. 11, 111601 (2016)
  doi:10.1103/PhysRevLett.117.111601
  [arXiv:1605.06098 [hep-th]].
  
  \bibitem{Das:2017pif} 
  S.~R.~Das, A.~Jevicki and K.~Suzuki,
  ``Three Dimensional View of the SYK/AdS Duality,''
  JHEP {\bf 1709}, 017 (2017)
  doi:10.1007/JHEP09(2017)017
  [arXiv:1704.07208 [hep-th]].
  
  \bibitem{Das:2017hrt} 
  S.~R.~Das, A.~Ghosh, A.~Jevicki and K.~Suzuki,
  ``Three Dimensional View of Arbitrary $q$ SYK models,''
  JHEP {\bf 1802}, 162 (2018)
  doi:10.1007/JHEP02(2018)162
  [arXiv:1711.09839 [hep-th]].

\bibitem{Das:2017wae} 
  S.~R.~Das, A.~Ghosh, A.~Jevicki and K.~Suzuki,
  ``Space-Time in the SYK Model,''
  JHEP {\bf 1807}, 184 (2018)
  doi:10.1007/JHEP07(2018)184
  [arXiv:1712.02725 [hep-th]].
  
  \bibitem{Kyono:2017jtc} 
  H.~Kyono, S.~Okumura and K.~Yoshida,
  ``Deformations of the Almheiri-Polchinski model,''
  JHEP {\bf 1703}, 173 (2017)
  doi:10.1007/JHEP03(2017)173
  [arXiv:1701.06340 [hep-th]].
  
  \bibitem{Kyono:2017pxs} 
  H.~Kyono, S.~Okumura and K.~Yoshida,
  ``Comments on 2D dilaton gravity system with a hyperbolic dilaton potential,''
  Nucl.\ Phys.\ B {\bf 923}, 126 (2017)
  doi:10.1016/j.nuclphysb.2017.07.013
  [arXiv:1704.07410 [hep-th]].
  
  \bibitem{Okumura:2018xbh} 
  S.~Okumura and K.~Yoshida,
  ``Weyl transformation and regular solutions in a deformed Jackiw-Teitelboim model,''
  Nucl.\ Phys.\ B {\bf 933}, 234 (2018)
  doi:10.1016/j.nuclphysb.2018.06.003
  [arXiv:1801.10537 [hep-th]].
  
  \bibitem{Jevicki:1980zg} 
  A.~Jevicki and B.~Sakita,
  ``Collective Field Approach to the Large $N$ Limit: Euclidean Field Theories,''
  Nucl.\ Phys.\ B {\bf 185}, 89 (1981)
  doi:10.1016/0550-3213(81)90365-5;~
  S.~R.~Das and A.~Jevicki,
  ``Large N collective fields and holography,''
  Phys.\ Rev.\ D {\bf 68}, 044011 (2003)
  doi:10.1103/PhysRevD.68.044011
  [hep-th/0304093].
  
 \bibitem{Klimcik:2002} C.~Klim\v{c}\'{i}k, ``Yang-Baxter sigma models and dS/AdS T duality,'' JHEP {\bf 0212}, 051 
(2002)  doi:10.1088/1126-6708/2002/12/051 [hep-th/0210095];  ~ C.~Klim\v{c}\'{i}k, ``On integrability of the Yang-Baxter 
sigma-model,'' J.\ Math.\ Phys.\  {\bf 50}, 043508 (2009) doi:10.1063/1.3116242 [arXiv:0802.3518 [hep-th]].
  
\bibitem{Delduc:2013} F.~Delduc, M.~Magro and B.~Vicedo, ``On classical $q$-deformations of integrable sigma-models,''
  JHEP {\bf 1311}, 192 (2013) doi:10.1007/JHEP11(2013)192 [arXiv:1308.3581 [hep-th]].  
  
\bibitem{Yoshida:2015a} T.~Matsumoto and K.~Yoshida, ``Yang-Baxter sigma models based on the CYBE,''
  Nucl.\ Phys.\ B {\bf 893}, 287 (2015) doi:10.1016/j.nuclphysb.2015.02.009 [arXiv:1501.03665 [hep-th]].
  
  \bibitem{Bakhmatov:2018apn} 
  T.~Araujo, I.~Bakhmatov, E.~Ó.~Colgáin, J.~Sakamoto, M.~M.~Sheikh-Jabbari and K.~Yoshida,
  ``Yang-Baxter $\sigma$-models, conformal twists, and noncommutative Yang-Mills theory,''
  Phys.\ Rev.\ D {\bf 95}, no. 10, 105006 (2017)
  doi:10.1103/PhysRevD.95.105006
  [arXiv:1702.02861 [hep-th]],~~
  I.~Bakhmatov, E.~Ó.~Colgáin, M.~M.~Sheikh-Jabbari and H.~Yavartanoo,
  ``Yang-Baxter Deformations Beyond Coset Spaces (a slick way to do TsT),''
  JHEP {\bf 1806}, 161 (2018)
  doi:10.1007/JHEP06(2018)161
  [arXiv:1803.07498 [hep-th]],~~T.~Kameyama and K.~Yoshida,
  ``Generalized quark-antiquark potentials from a $q$-deformed AdS$_5 \times $S$^5$ background,''
  PTEP {\bf 2016}, no. 6, 063B01 (2016)
  doi:10.1093/ptep/ptw059
  [arXiv:1602.06786 [hep-th]],~~T.~Araujo, I.~Bakhmatov, E.~Ó.~Colgáin, J.~i.~Sakamoto, M.~M.~Sheikh-Jabbari and K.~Yoshida,
  ``Conformal twists, Yang-Baxter $ \sigma $-models \& holographic noncommutativity,''
  J.\ Phys.\ A {\bf 51}, no. 23, 235401 (2018)
  doi:10.1088/1751-8121/aac195
  [arXiv:1705.02063 [hep-th]].
  
\bibitem{Beisert:2012} N.~Beisert, W.~Galleas and T.~Matsumoto, ``A Quantum Affine Algebra for the Deformed Hubbard 
Chain,'' J.\ Phys.\ A {\bf 45} (2012) 365206 doi:10.1088/1751-8113/45/36/365206 [arXiv:1102.5700 [math-ph]].

\bibitem{Minahan:2002} J.~A.~Minahan and K.~Zarembo, ``The Bethe ansatz for N=4 superYang-Mills,''
  JHEP {\bf 0303} (2003) 013 doi:10.1088/1126-6708/2003/03/013 [hep-th/0212208].  
  
\bibitem{Yoshida:2014} I.~Kawaguchi, T.~Matsumoto and K.~Yoshida, ``Jordanian deformations of the $AdS_5\times S^5$ 
superstring,'' JHEP {\bf 1404} (2014) 153 doi:10.1007/JHEP04(2014)153 [arXiv:1401.4855 [hep-th]].  
  
\bibitem{Delduc:2013a} F.~Delduc, M.~Magro and B.~Vicedo, ``An integrable deformation of the $AdS_5\times S^5$ superstring 
action,'' Phys.\ Rev.\ Lett.\  {\bf 112}, no. 5, 051601 (2014) doi:10.1103/PhysRevLett.112.051601 [arXiv:1309.5850 [hep-th]];~ 
F.~Delduc, M.~Magro and B.~Vicedo, ``Derivation of the action and symmetries of the $q$-deformed $AdS_{5} \times S^{5}$ 
superstring,'' JHEP {\bf 1410}, 132 (2014) doi:10.1007/JHEP10(2014)132 [arXiv:1406.6286 [hep-th]].  
 
  \bibitem{Kameyama:2014vma} 
  T.~Kameyama and K.~Yoshida,
  ``A new coordinate system for $q$-deformed AdS$_{5} \times$ S$^5$ and classical string solutions,''
  J.\ Phys.\ A {\bf 48}, no. 7, 075401 (2015)
  doi:10.1088/1751-8113/48/7/075401
  [arXiv:1408.2189 [hep-th]];~~T.~Kameyama and K.~Yoshida,
  ``Minimal surfaces in $q$-deformed AdS$_5\times$S$^5$ with Poincare coordinates,''
  J.\ Phys.\ A {\bf 48}, no. 24, 245401 (2015)
  doi:10.1088/1751-8113/48/24/245401
  [arXiv:1410.5544 [hep-th]];~~D.~Roychowdhury,
  ``Stringy correlations on deformed $ AdS_{3}\times S^{3} $,''
  JHEP {\bf 1703}, 043 (2017)
  doi:10.1007/JHEP03(2017)043
  [arXiv:1702.01405 [hep-th]].


\end{thebibliography}
\end{document}